\documentclass[preprint, 12pt]{elsarticle}

\usepackage{hyperref}
\usepackage{lineno}
\modulolinenumbers[0]

\usepackage{amsmath,amssymb}
\usepackage{siunitx}
\DeclareSIUnit\speedoflight{c}
\DeclareSIUnit\rad{rad}

\usepackage[T1]{fontenc}
\usepackage[utf8]{inputenc}
\usepackage{xspace}

\usepackage{tikz}
\usetikzlibrary{%
  arrows,
  shadows,
  calc,
  patterns,
  positioning,
  decorations.pathreplacing,
  decorations.markings,
  trees
}
\tikzset{>=latex}

\usepackage{listings}%

\usepackage{subcaption}

\usepackage{booktabs}

\journal{Computer Physics Communications}

\newcommand{\tabref}[1]{Table~\ref{#1}}
\newcommand{\figref}[1]{Fig.~\ref{#1}}
\newcommand{\lstref}[1]{Listing~\ref{#1}}

\newcommand{\moliere}{Moli\`{e}re\xspace}
\newcommand{\frejus}{Fr\'{e}jus\xspace}

\newcommand{\Li}{\ensuremath{\operatorname{Li}}}
\begin{document}

\begin{frontmatter}

\title{Recent Improvements for the Lepton Propagator PROPOSAL}
\author[TUDo]{Mario Dunsch}
\author[TUDo]{Jan Soedingrekso}
\author[TUDo]{Alexander Sandrock}
\author[TUDo]{Maximilian Meier}
\author[TUDo]{Thorben Menne}
\author[TUDo]{Wolfgang Rhode}
\address[TUDo]{Department of Physics, TU Dortmund University, D-44221 Dortmund,
 Germany}

\begin{abstract}
The lepton propagator PROPOSAL is a Monte-Carlo Simulation library written in C++,
propagating high energy muons and other charged particles through large distances of media.
In this article, a restructuring of the code is described,
which yields a performance improvement of up to \SI{30}{\percent}.
For an improved accuracy of the propagation processes,
more exact calculations of the leptonic and hadronic decay process and
more precise parametrizations for the interaction cross sections are now available.
The new modular structure allows a more flexible and custom usage,
which is further facilitated with a python interface.
\end{abstract}


\end{frontmatter}
\section{Introduction}
Very large volume neutrino telescopes such as IceCube, ANTARES, GVD and the
planned facilities IceCube-Gen2 and KM3net play a key role in searches for
astrophysical neutrinos, the origin of cosmic rays, dark matter and exotic
relic particles expected from the early universe. In all these
investigations, the key process is the detection of light emitted by particles
propagating through the detector and the adjacent medium. These particles,
with energies from \si{\giga\electronvolt} up to several \si{\peta\electronvolt} or 
even \si{\exa\electronvolt}, are charged leptons produced in extended air showers or
neutrino interactions and possibly also heavy exotic particles.

The propagation of leptons through matter is a task, for which several codes
have appeared over the years \cite{music,mum,geant1,geant2,geant3,mmc,%
PROPOSAL}. This article reports a major update to PROPOSAL\footnote{
The code is available at
\url{https://github.com/tudo-astroparticlephysics/PROPOSAL}.}
\cite{PROPOSAL}, the \textbf{Pr}opagator with \textbf{o}ptimal
\textbf{p}recision and \textbf{o}ptimized \textbf{s}peed for \textbf{a}ll
\textbf{l}eptons.
The key requirements that a lepton propagator has to fulfill are a physical
description that is as accurate as possible, while reducing computational
errors and runtime as much as possible. To improve the accuracy of the
propagation processes, precise parametrizations of the cross sections and
a detailed treatment of decay processes are required.

The current version of PROPOSAL implements most recent cross section
parametrizations and improves the treatment of both leptonic and hadronic
decays by a more exact description of final state kinematics.
The code has been restructured, which resulted in a
significant increase of execution speed. The description of particles and their
interaction has been changed, making heavy use of polymorphism, which
facilitates the extension of PROPOSAL for the propagation of other particles;
as an example, the treatment of sleptons predicted by supersymmetry is shown.
Finally, a Python interface was added to allow easier usage and to enable
the use of PROPOSAL in Python.

\section{Updates}
\label{sec:neuerungen}

\subsection{Updates of the code structure}
\label{sub:neuerungen_der_code_struktur}

% \begin{itemize}
%   \item Polymorphic class structure \\
%     $\Rightarrow$ Easy to extend
%     $\Rightarrow$ New particles can be constructed without modifying
%     the source code
%   \item \texttt{Python}-interface \\
%     $\Rightarrow$ Easy operation for the user
%     (code-snippets with explanations)
%   \item Performance improvements
%     (comparison with the old code, inclusively the performance of the
%     different scattering models)
% \end{itemize}

The Software PROPOSAL is a further development of the former
program MMC (\textbf{M}uon \textbf{M}onte \textbf{C}arlo) \cite{mmc},
which was written in Java where the version dependency of the code became
a huge drawback. Therefore, PROPOSAL was developed in C++ based on MMC
providing the same precision and increased performance \cite{PROPOSAL}.
In the first version of PROPOSAL, the code structure of MMC was reproduced.
This code was now restructured to fit a more modern object-oriented C++ coding style.
In particular, the data needed for the propagation routines are stored in
corresponding classes, and polymorphism
is used to get rid of the reliance on runtime type information.

%%%%%%%%%%%%%%%%%%%%%%%%%%%%%%%%%%%%%%%%%%%%%%%%%%%%%%%%%%%%%%%%%%%%%%%%%%%%%%%
%                    explanation of constructor arguments                     %
%%%%%%%%%%%%%%%%%%%%%%%%%%%%%%%%%%%%%%%%%%%%%%%%%%%%%%%%%%%%%%%%%%%%%%%%%%%%%%%

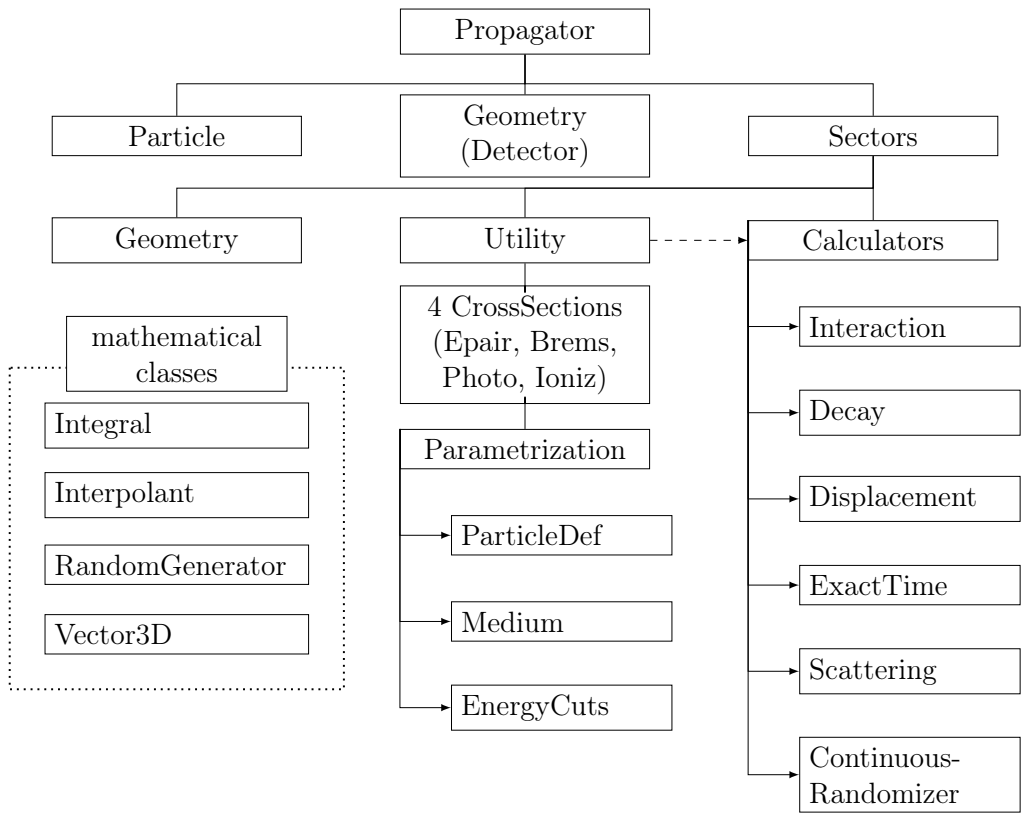
\begin{figure}[htb]
  \centering
  \resizebox {\textwidth} {!} {
  \begin{tikzpicture}[
    sibling distance=50mm, >=latex,
    edge from parent fork down,
    basic/.style  = {draw, text width=8em, rectangle},
    root/.style   = {basic, align=center},
    level 2/.style = {basic, align=center, text width=8em},
    level 3/.style = {basic, thin, align=center, text width=8em},
    level 4/.style = {basic, thin, align=left, text width=7em}
  ]

  % root of the the initial tree, level 1
  \node[root] {Propagator}
  % The first level, as children of the initial tree
    child {node[level 2] (c1) {Particle}}
    child {node[level 2] (c2) {Geometry (Detector)}}
    child {node[level 2] (c3) {Sectors}
      child {node[level 3] (d1) {Geometry}}
      child {%
        node[level 3] (d2) {Utility}
          child {%
            node[level 3] (f1) {4 CrossSections (Epair, Brems, Photo, Ioniz)}
            child {%
              node[level 3] (e1) {Parametrization}
            }
          }
        }
      child {node[level 3] (d3) {Calculators}}
      child [missing]
      child [missing]};

  \node [basic, text width=8.5em, below = 2cm of d1] (m1) {Integral};
  \node [basic, text width=8.5em, below of = m1] (m2) {Interpolant};
  \node [basic, text width=8.5em, below of = m2] (m3) {RandomGenerator};
  \node [basic, text width=8.5em, below of = m3] (m4) {{Vector3D}};

  % The second level, relatively positioned nodes
  \begin{scope}[every node/.style={level 4, node distance=3em}]

  \node [below of = e1, xshift=15pt] (d21) {ParticleDef};
  \node [below of = d21] (d22) {Medium};
  \node [below of = d22] (d23) {EnergyCuts};

  \node [below of = d3, xshift=15pt] (d31) {Interaction};
  \node [below of = d31] (d32) {Decay};
  \node [below of = d32] (d33) {Displacement};
  \node [below of = d33] (d34) {ExactTime};
  \node [below of = d34] (d35) {Scattering};
  \node [node distance=3.6em, below of = d35] (d36) {Continuous- Randomizer};
  \end{scope}

  \foreach \value in {1,...,3}
    \draw[->] (e1.north west) |- (d2\value.west);

  \foreach \value in {1,...,6}
    \draw[->] (d3.north west) |- (d3\value.west);

  \draw[->, dashed] (d2.east) -- (d3.west);

  \draw[thick,dotted] ($(m1.north west)+(-0.5,0.5)$) rectangle ($(m4.south east)+(0.5,-0.5)$);
  \node[draw, fill=white, rectangle, text width=7em, align=center] at ($(m1.north) + (0, 0.7)$) {mathematical classes};

  \end{tikzpicture}
  }
  \caption{ Schematic of the main class structure.}

  \label{fig:schematic_propagator}
\end{figure}

The basic structure of the current code is shown
in \figref{fig:schematic_propagator}.
Here the base class of PROPOSAL is the \texttt{Propagator} class, which
holds the particle to be propagated, a geometry describing the detector volume
and a list of at least one sector through which the particle is to be propagated.
These parts are set once with the construction of the propagator and
cannot be changed afterwards. The constructor takes a further argument,
the \texttt{InterpolationDef} (\lstref{lst:constructor}), which is used to
determine whether to integrate or interpolate the implemented cross sections.
If the interpolation is chosen, PROPOSAL saves interpolation tables in
memory or on disk as determined by the \texttt{InterpolationDef}.
These tables are sensitive to the parameters of the particle, sectors and
especially the cross sections.
Therefore the API prevents the user from changing the
parameters of the propagator after the initialization, because the new state will not match with the generated tables.
If the user needs to change the propagator, he is forced to create a new one.

In the following paragraph, the remaining three arguments of the
constructor will be discussed.
The first argument of the propagator is the \texttt{ParticleDef},
holding static data of the particle like the mass, charge, lifetime, decay modes
and $e_{low}$, defining the energy below which the particle is treated as lost
(defaults to the mass of the particle). The propagator creates
an instance of the \texttt{Particle} class, a dynamic particle,
out of the \texttt{ParticleDef} which is a composition of the static
particle definition and member variables like position, energy, and momentum.
The user can get a reference to the particle via a getter method to read
out the information. However, the particle definition should not be changed,
therefore all the members of \texttt{ParticleDef} are declared as
\lstinline[language=C++]{const}.
There are several predefined particle definitions like the
\texttt{MuMinusDef} or the \texttt{TauMinusDef} which are derived from
the \texttt{ParticleDef} and obtained as singletons with
\texttt{MuMinusDef::Get()}.
To define custom particle definitions there is a
\texttt{ParticleDef::Builder} class provided, which allows, for example, the
creation of muon definitions with different masses directly from
the \texttt{MuMinusDef}.
\begin{lstlisting}[
  float=ht,
  numbers=left,
  numberstyle=\tiny,
  framexleftmargin=1.5em,
  xleftmargin=2em,
  frame=single,
  caption=Constructor of the propagator class,
  label={lst:constructor},
  language=C++
]
  Propagator(const ParticleDef&,
             const std::vector<Sector::Definition>&,
             const Geometry&,
             const InterpolationDef&)
\end{lstlisting}
The core of the Propagator is defined by a list of sector definitions.
In \tabref{tab:sector_def} the individual parameters are shown along with
a short description.
Noteworthy is the choice of a model for multiple scattering.
As opposed to the previous version of PROPOSAL, two further multiple scattering models were added,
which will be discussed in section \ref{sub:physikalische_neuerungen}.
Furthermore, there is another definition object, the \texttt{utility\_def},
which contains the definition of the individual cross section parameters for
the bremsstrahlung, pair production, photonuclear interaction, and ionization.
These parameters include a multiplier to manually scale the cross sections,
a \lstinline[language=C++]{bool} to decide whether to consider the LPM-effect (bremsstrahlung and
pair production only) and an \lstinline[language=C++]{enum}
for the choice of the parametrization
(bremsstrahlung and photonuclear interaction only).
\begin{table}[htpb]
  \caption{Description of the parameters for the sector definitions.}
  \centering
  \begin{tabular}{lp{8cm}}
    \toprule
    Parameter & Description \\
    \midrule
    \texttt{Medium} & Medium of the sector \\
    \texttt{EnergyCutsSettings} & Stores $e_\text{cut}$ and $v_\text{cut}$ \\
    \texttt{Geometry} & Geometry of the sector \\
    \texttt{stopping\_decay} & Whether to force a final decay of the particle if its energy is $\leq e_\text{cut}$ \\
    \texttt{cont\_rand} & Whether to use continuous randomization \\
    \texttt{exact\_time} & Whether to calculation the time exactly out of the tracking integral or to use an approximation\\
    \texttt{scattering\_model} & Choice of the multiple scattering model, \texttt{HighlandIntegral}, \texttt{Highland} or \texttt{Moliere} \\
    \texttt{particle\_location} & Location of the particle \\
    \texttt{utility\_def} & Definition of cross section parameters \\
    \bottomrule
  \end{tabular}
  \label{tab:sector_def}
\end{table}
%
%%%%%%%%%%%%%%%%%%%%%%%%%%%%%%%%%%%%%%%%%%%%%%%%%%%%%%%%%%%%%%%%%%%%%%%%%%%%%%%
%                             Maybe include this                              %
%%%%%%%%%%%%%%%%%%%%%%%%%%%%%%%%%%%%%%%%%%%%%%%%%%%%%%%%%%%%%%%%%%%%%%%%%%%%%%%

The main routine of the Propagator is the \texttt{Propagate} method shown
in \lstref{lst:propagate}. This method propagates the particle through
the previously defined sectors and returns a list of secondaries
expressed as \texttt{DynamicData}.
These secondaries can be, determined by the id of this class, stochastic
energy losses or particles in case of a particle decay.
\begin{lstlisting}[
  float=ht,
  numbers=left,
  numberstyle=\tiny,
  framexleftmargin=1.5em,
  xleftmargin=2em,
  frame=single,
  caption=Constructor of the propagator class,
  label={lst:propagate},
  language=C++
]
  std::vector<DynamicData*>
  Propagate(double MaxDistance_cm = 1e20);
\end{lstlisting}
%
%%%%%%%%%%%%%%%%%%%%%%%%%%%%%%%%%%%%%%%%%%%%%%%%%%%%%%%%%%%%%%%%%%%%%%%%%%%%%%%
%                              End maybe include                              %
%%%%%%%%%%%%%%%%%%%%%%%%%%%%%%%%%%%%%%%%%%%%%%%%%%%%%%%%%%%%%%%%%%%%%%%%%%%%%%%
%
%
%%%%%%%%%%%%%%%%%%%%%%%%%%%%%%%%%%%%%%%%%%%%%%%%%%%%%%%%%%%%%%%%%%%%%%%%%%%%%%%
%                              Python interface                               %
%%%%%%%%%%%%%%%%%%%%%%%%%%%%%%%%%%%%%%%%%%%%%%%%%%%%%%%%%%%%%%%%%%%%%%%%%%%%%%%

As the programming language \texttt{Python} is getting more and more
popular in scientific applications, a \texttt{Python}-interface,
which can be created as a build option of PROPOSAL, is now provided.
The library \texttt{Boost.Python} \cite{Boost} is used for the interface.
A typical \texttt{Python} code for PROPOSAL is shown in \lstref{lst:python}.
The comments in the listing explain some details of the different parts.
% , but with the descriptions given above the code is written to be self-explanatory.
In this example, a propagator is created with only one sector and $10^6$ muons
are propagated through this sector to obtain a list of muon ranges.
\begin{lstlisting}[
  float=htp,
  basicstyle=\scriptsize,
  numbers=left,
  numberstyle=\tiny,
  framexleftmargin=3em,
  xleftmargin=2.5em,
  frame=single,
  caption=\texttt{Python} code listing showing the basic instantiation of
  the propagator and creating data to visualize the muon ranges.,
  label={lst:python},
  language=python
]
  import pyPROPOSAL as pp

  sec_def = pp.SectorDefinition()
  sec_def.medium = pp.medium.Ice(density_correction=0.98)
  # Init Sphere(position, outer radius, inner radius)
  sec_def.geometry_def = pp.geometry.Sphere(pp.Vector3D(), 1e20, 0)

  sec_def.scattering_model = pp.scattering.ScatteringModel.Moliere
  sec_def.crosssection_defs.brems_def.lpm_effect = False
  sec_def.crosssection_defs.epair_def.lpm_effect = False

  sec_def.cut_settings.ecut = 500
  sec_def.cut_settings.vcut = 0.05

  interpolation_def = pp.InterpolationDef()
  interpolation_def.path_to_tables = "~/.local/share/PROPOSAL"

  prop = pp.Propagator(
          particle_def=pp.particle.MuMinusDef.get(),
          sector_defs=[sec_def],
          detector=pp.geometry.Sphere(pp.Vector3D(), 1e20, 0),
          interpolation_def=interpolation_def
  )

  mu = prop.particle

  mu_length = []

  for i in range(int(1e6)):
      # Initial settings of the muon
      mu.position = pp.Vector3D(0, 0, 0)
      mu.direction = pp.Vector3D(0, 0, -1)
      mu.energy = 1e6  # MeV
      mu.propagated_distance = 0  # cm

      secondaries = prop.propagate()

      mu_length.append(mu.propagated_distance / 100)
\end{lstlisting}
%
%%%%%%%%%%%%%%%%%%%%%%%%%%%%%%%%%%%%%%%%%%%%%%%%%%%%%%%%%%%%%%%%%%%%%%%%%%%%%%%
%                                 performance                                 %
%%%%%%%%%%%%%%%%%%%%%%%%%%%%%%%%%%%%%%%%%%%%%%%%%%%%%%%%%%%%%%%%%%%%%%%%%%%%%%%

\begin{figure}[htb]
  \centering
  \includegraphics[scale=1.0]{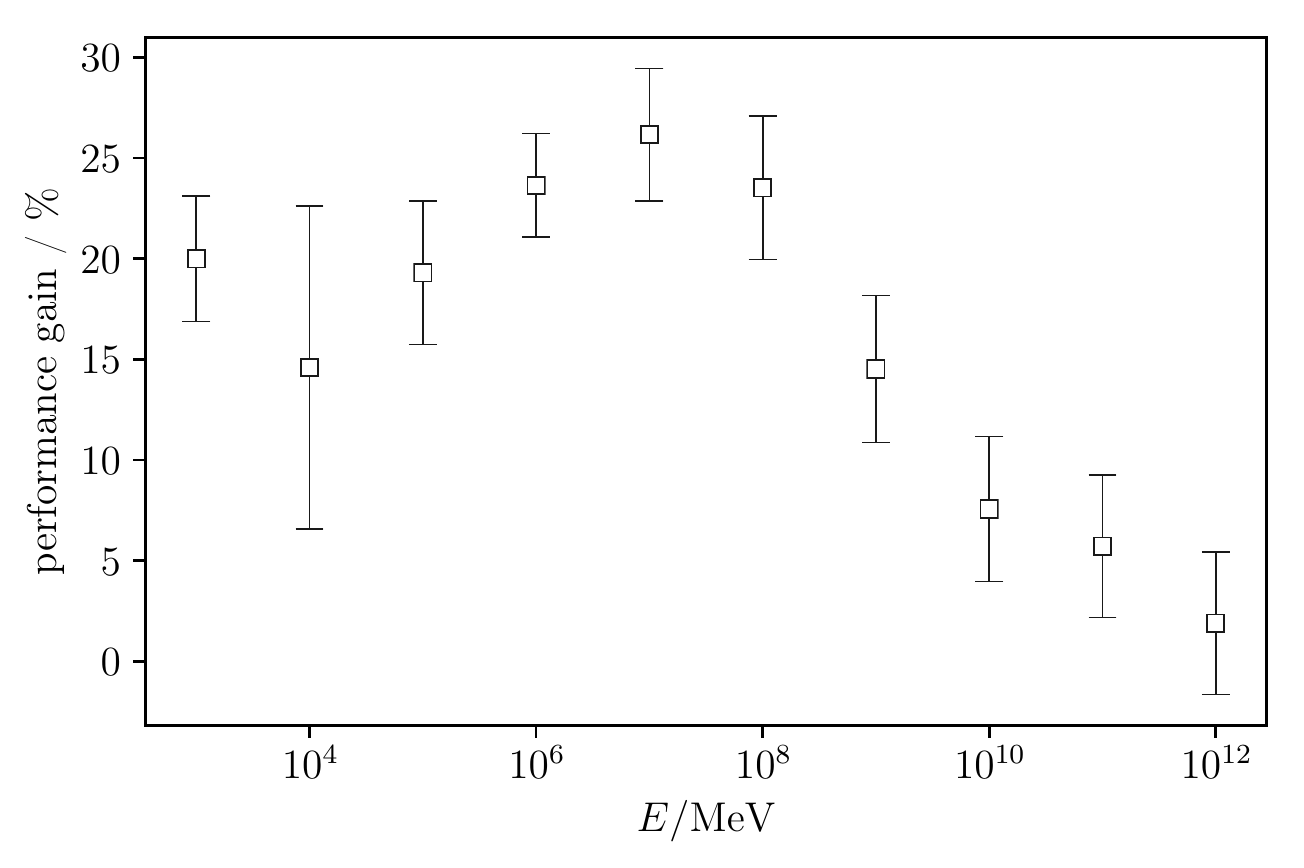}
  \caption{Runtime improvement $(t_\text{old} - t_\text{new})/t_\text{old}$
  of the new version compared to the previous version. Multiple scattering is disabled.
  Per energy range $1000$ muons were propagated through ice until they lost their
  energy.}
\label{fig:performace_wo_scattering}
\end{figure}

The new code restructuring also has the effect of a performance gain.
Previously the transition from MMC to PROPOSAL already came with a
performance gain of up to \SI{40}{\%} \cite{JHK}.
This test was reproduced with the actual version of PROPOSAL and
without multiple scattering to measure only the core routines.
The result is shown in \figref{fig:performace_wo_scattering}.
Especially in the energy range relevant for the high energy physics,
a performance gain of up to \SI{25}{\%} is obtained.
A laptop computer with an
\textit{Intel\textsuperscript{~\textregistered} Core\textsuperscript{\texttrademark} i5-4200U}
processor was used for these benchmarks.

\newpage
\subsection{Physical updates}
\label{sub:physikalische_neuerungen}

The physical improvements in the current version can be categorised in two parts; a whole restructuring of the decay process, especially for the hadronic $\tau$ decay and a broadening repertory of more accurate cross sections and multiple scattering parametrizations.

\subsubsection{New Decay Implementation}
The decay routines in PROPOSAL are divided into leptonic decays and hadronic decays, as already described in \cite{PROPOSAL}.
\paragraph{Leptonic Decay}
For the leptonic decays, the energy distribution of the produced leptons are calculated with the differential decay width in the rest frame of the decaying lepton \cite{pdg18} % eq. 63.3
\begin{align}
  \frac{\mathrm{d}\Gamma}{\mathrm{d}x} =
  \frac{G_\text{F}^2 M^5}{192 \pi^3} (3 - 2x)x^2,
  	\quad x = \frac{E_l}{E_{\mathrm{max}}}
\end{align}
with the Fermi constant $G_\text{F}$, the mass of the decaying lepton $M$ and the limits for the energy of the produced leptons $E_l$ from their mass $m_l$ to $E_{\mathrm{max}} = (M^2 + m_l^2)/2M$.

In this parametrization of the decay width, the approximation $m_l^2 / M^2 \approx 0$ was applied, which is good for the muon decay ($m_{e} / m_{\mu} \approx 1/200$) and even better for the electronic tau decay. However, for the muonic tau decay the mass ratio ($m_{\mu} / m_{\tau} \approx 1/17$) is not small and the approximation is not valid anymore.
Therefore a differential decay width without this approximation (e.\,g. \cite{LahiriPal}) is used
\begin{align}
  \frac{\mathrm{d}\Gamma}{\mathrm{d}x} =
    \frac{G_\text{F}^2}{12 \pi^3} E_{\mathrm{max}} \sqrt{E_l^2 - m_l^2}
    \left[ M E_l(3 M - 4 E_l) + m_l^2 (3 E_l - 2 M) \right] .
\end{align}
The integrated expression is, for the approximate decay width, just a polynomial ($x^3(1 - x/2)$), while, for the more accurate decay width, it is a more complex expression
\begin{align}
  \int \frac{\mathrm{d}\Gamma}{\mathrm{d}x} \mathrm{d}x &=
    \frac{G_\text{F}^2}{12 \pi^3} \left( \frac{3}{2} m_l^4 M \log(\sqrt{E_l^2 - m_l^2} + E_l) \right. \\
    & + \left. \sqrt{E_l^2 - m_l^2} [ (M^2 + m_l^2 - M E_l) (E_l^2 - m_l^2) - 1.5 M E_l m_l^2] \right) .
\end{align}
Since both integrals are not invertible, a root finding algorithm (Newton-Raphson method from \cite{Boost}) is used to transform the uniformly sampled random numbers into the desired form.
Therefrom the expressions are called multiple times and the higher accuracy of this parametrization goes along with a slower sampling (6 times slower) from this distribution.
% [Soll ich noch sagen, dass der user dazwischen auswählen können soll, oder nicht? Können es ja so machen, wie mit dem Raubold Lynch Sampling mit Matrixelementen (Gewichten) oder dem Resonanz-Zweikörperzerfall]

\begin{figure}
    \centering
    \includegraphics[width=\textwidth]{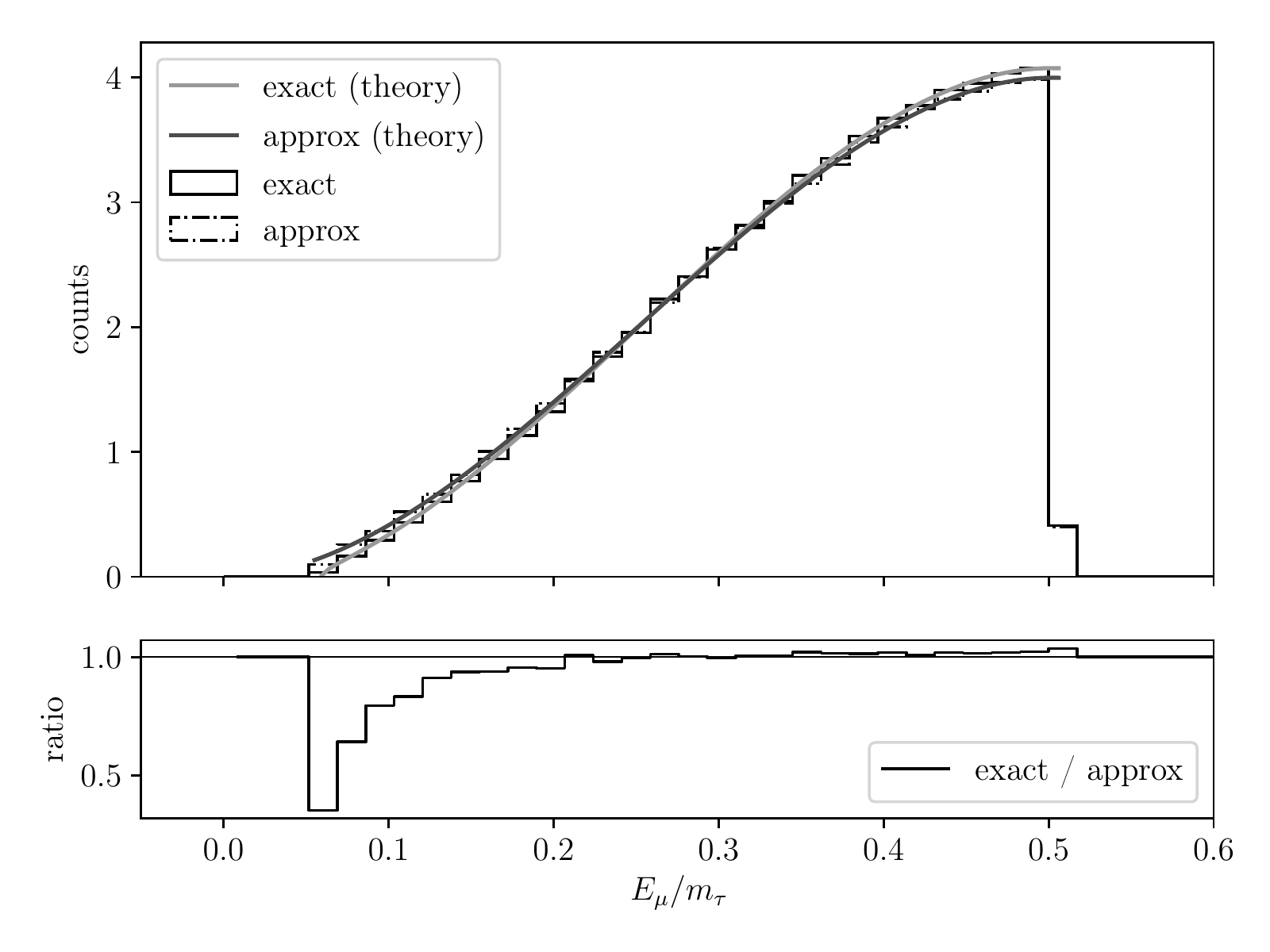}
    \caption{Normalized energy distribution of the muons in the rest frame of a
      muonic tau decay. Shown are the distribution with and without the
      approximation $m_{\mu}^2 / m_{\tau}^2 \approx 0$.
      The binned data are the results of $10^6$ simulated tau decays $\tau
      \rightarrow \mu \nu \nu$ while the solid lines represents the
      theoretical predictions.}
    \label{fig:tau_mu_decay_distribution}
\end{figure}

The comparison between these two distributions is shown in \figref{fig:tau_mu_decay_distribution}.

\begin{figure}[htb]
    \centering
    \includegraphics[width=\textwidth]{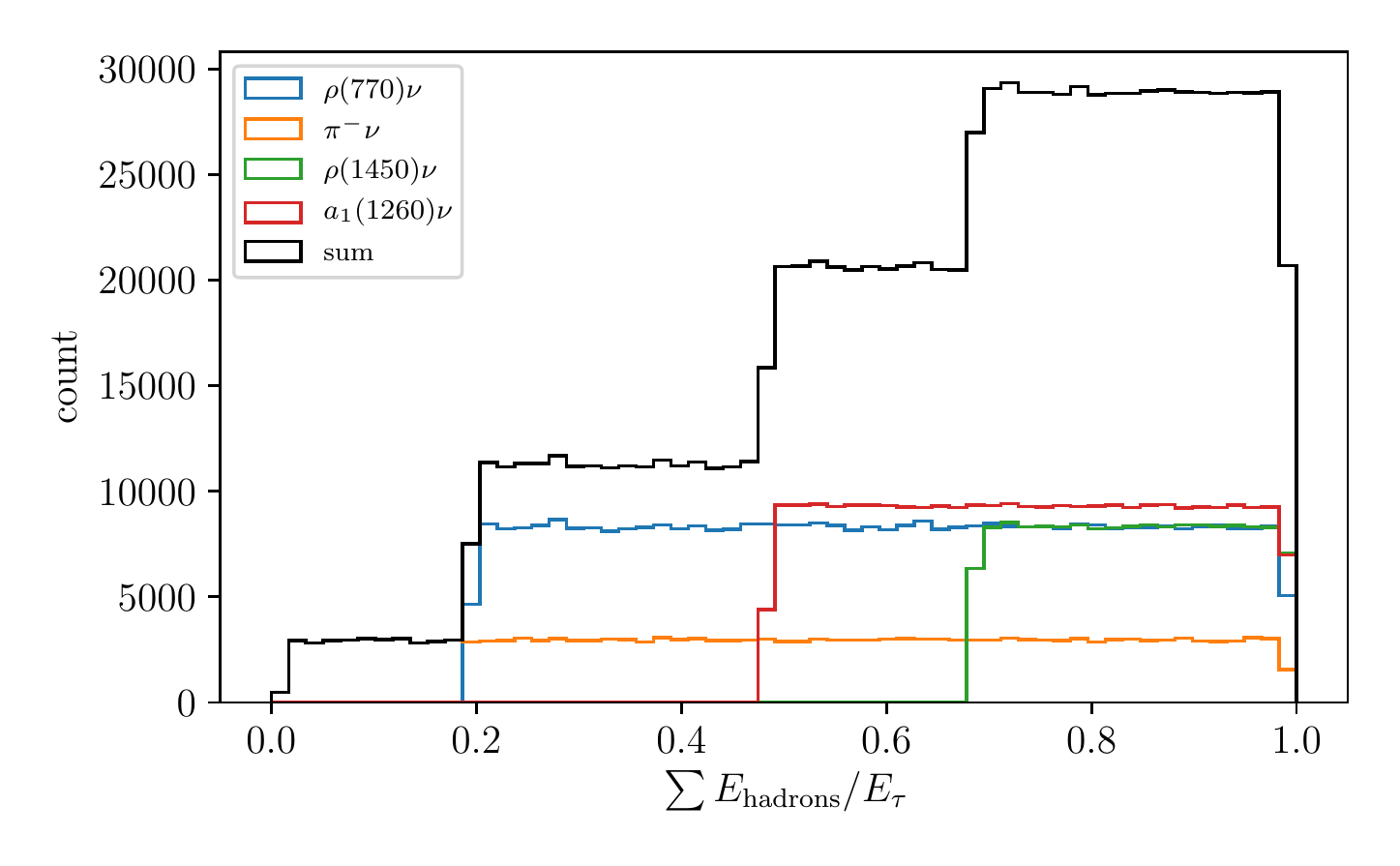}
    \caption{Energy distribution of the secondaries in the laboratory frame.}
    \label{fig:decay_dist_two_body}
\end{figure}

\paragraph{Hadronic Decay}
In the old version, a two body decay approximation was used
for the hadronic decay \cite{PROPOSAL}, while leaving the matrix element constant and set to one.
The hadronic secondaries produced are represented by a heavier meson or resonances,
which mainly decay into the desired hadronic secondaries. The decay modes are
listed in \tabref{tab:two_body_decay_channels}. The energy of each
resonance with mass $m_R$ in the rest frame of the decaying lepton is
$E_\mathrm{rest} = (M^2 + m_l^2)/2M$, which can be identified as peaks in
the energy distribution of the hadronic secondaries. After boosting in the
laboratory system, these peaks produce steps in the secondary energy
distribution each time a new particle mass is reached, as can be seen in
\figref{fig:decay_dist_two_body}.
\begin{table}[htb]
  \centering
  \caption{Decay modes of the tau lepton used in the previous version \cite{mmc}.}
  \begin{tabular}{lc}
    \toprule
    Decay mode & Branching ratio / $\si{\%}$ \\
    \midrule
    $\mu$ & $17.37$ \\
    $e$ & $17.83$ \\
    \midrule
    $\pi$ & $11.09$ \\
    $\rho-770$ & $25.40$ \\
    $a_1-1260$ & $18.26$ \\
    $\rho-1465$ & $10.05$ \\
    \bottomrule
  \end{tabular}
  \label{tab:two_body_decay_channels}
\end{table}

In the new version, hadronic decays are treated as $n$-body decays, increasing the phase space
and smoothen the secondary distribution. For the sampling in this phase space
the Raubold-Lynch algorithm \cite{Byckling} is used. The idea behind this
algorithm is that the $n$-body phase space is recursively split up into $n$ 2-body
phase spaces.
%, as shown in figure \ref{fig:raubold_lynch}.
%Regarding a hypothetical decay into three massless particles, a quick check, if the algorithm was implemented correctly, was performed.
%The phase space volume in this case simplifies to $R_3 = M^2\pi/8$ and for the tau mass the volume is \SI{3.8913086}{\square\giga\electronvolt}.
%When sampling $10^6$ points in the phase space for a $\tau$ decay, the integration results in a phase space of \SI{3.8913\pm0.0018}{\square\giga\electronvolt}, which agrees with the theoretical prediction.

With this algorithm, more decay channels can be implemented, which can be seen in \tabref{tab:decay_modes}.
\begin{table}[htb]
    \centering
    \caption{Hadronic decay modes of the tau lepton with the highest branching ratios \cite{pdg18}.}
    \label{tab:decay_modes}
    \begin{tabular}{l S[table-format=2.4]}
        \toprule
        Decay mode & {Branching ratio / $\si{\%}$} \\
        \midrule
        $\pi^-$              & 10.82 \\
        $K^-$                & 0.70 \\
        $\pi^- \pi^0$        & 25.49 \\
        $K^- \pi^0$          & 0.43 \\
        $\pi^- 2\pi^0$       & 9.26 \\
        $\pi^- 3\pi^0$       & 1.04 \\
        $\pi^- K^0$          & 0.83 \\
        $\pi^- \pi^0 K^0$    & 0.38 \\
        $\pi^- \pi^- \pi^+$  & 8.99 \\
        $\pi^- \pi^- \pi^+ \pi^0$ & 2.74 \\
        $\pi^- \omega$       & 1.95 \\
        $\pi^- \pi^+ K^-$    & 0.29 \\
        \bottomrule
    \end{tabular} % sum is 98.1642, thats to far away from 100%
\end{table}

\begin{figure}[htpb]
  \centering
  \begin{subfigure}[b]{\textwidth}
    \includegraphics[width=\textwidth]{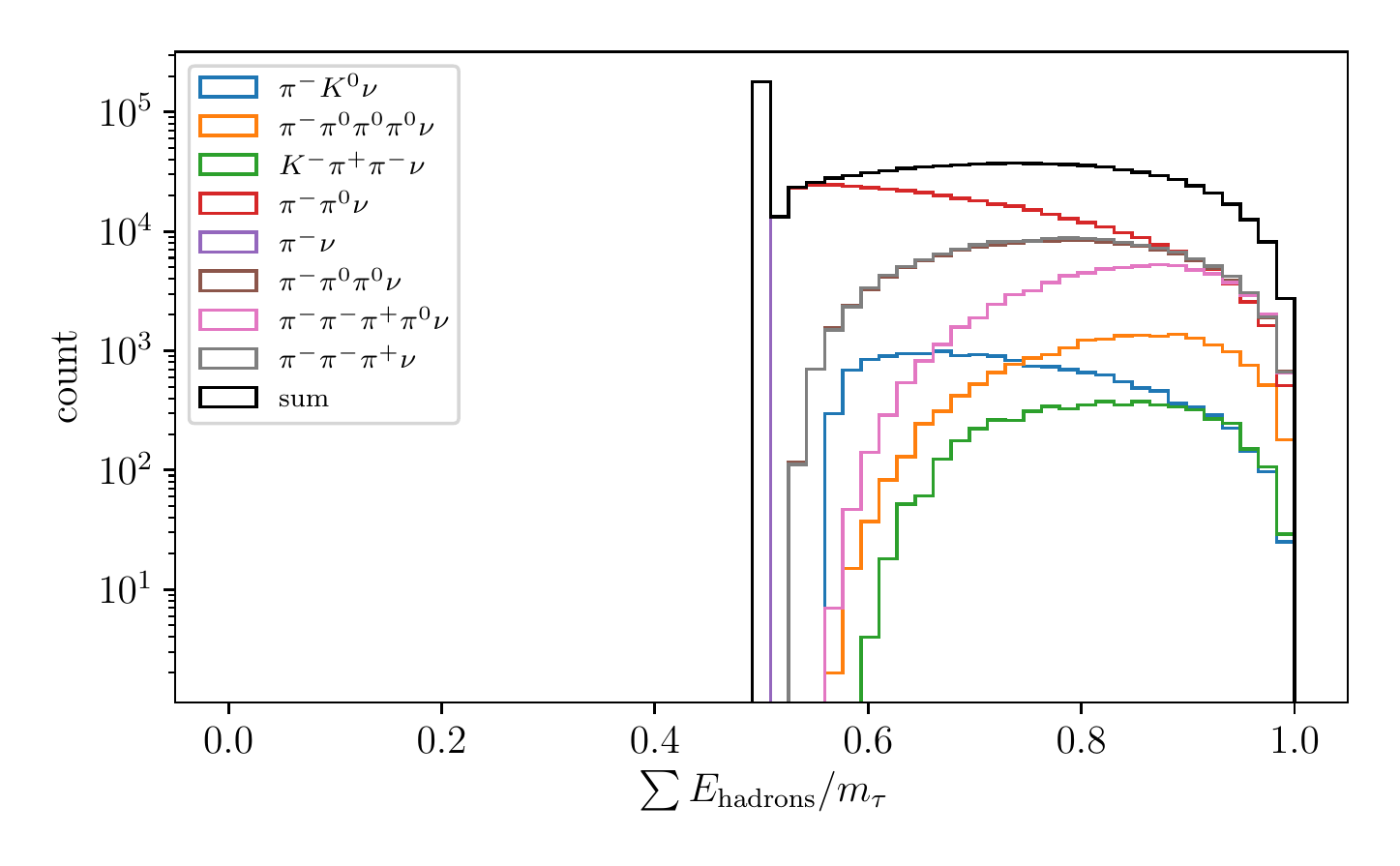}
    \caption{rest frame}
    \label{fig:hdr_dist}
  \end{subfigure}
  \begin{subfigure}[b]{\textwidth}
    \includegraphics[width=\textwidth]{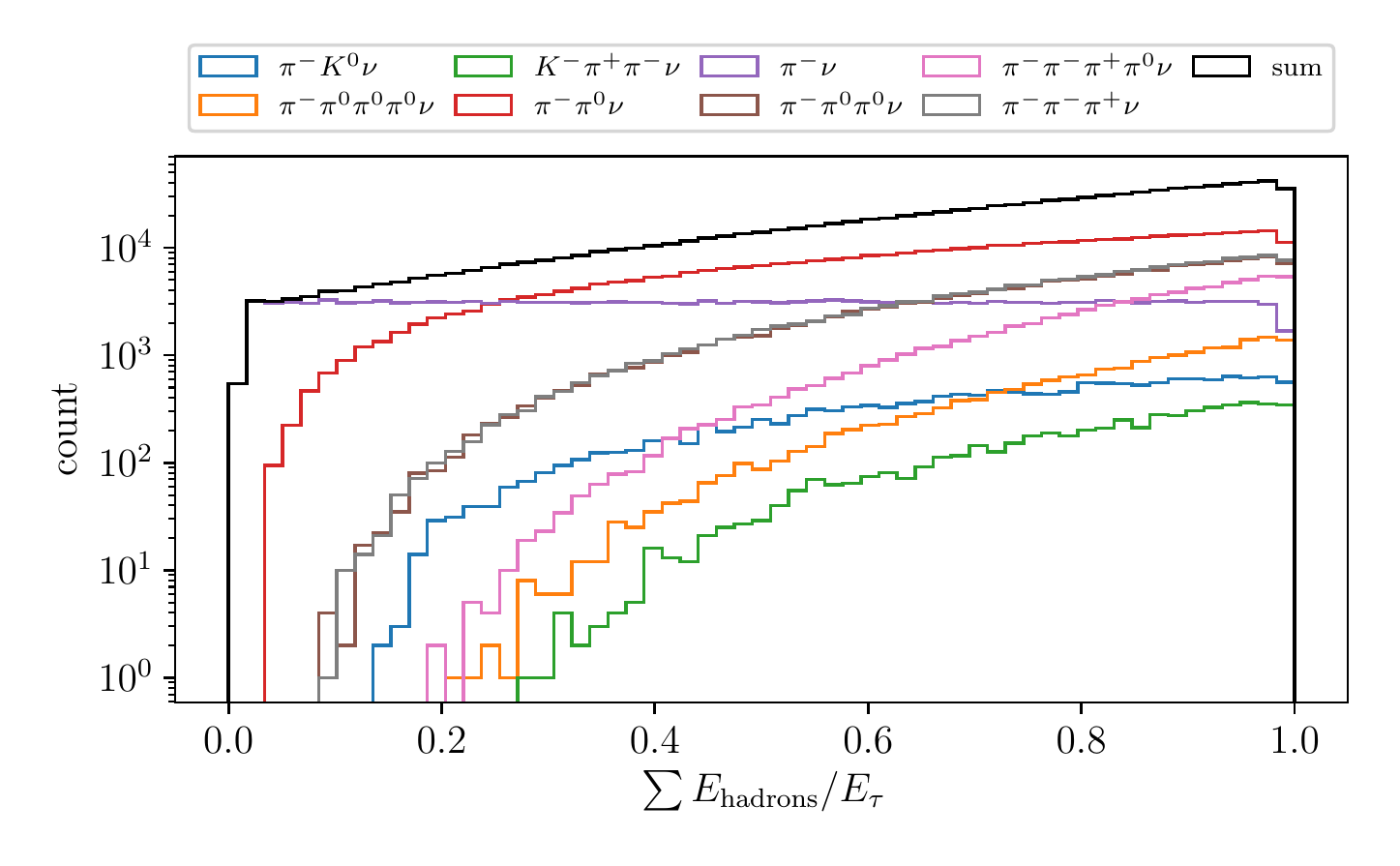}
    \caption{laboratory frame}
    \label{fig:hdr_dist_uniform}
  \end{subfigure}
  \caption{Energy distribution of the secondaries in hadronic tau decays
    in the rest frame (not boosted) and in the laboratory frame (boosted).}
  \label{fig:decay_dist_n_body}
\end{figure}

\begin{figure}[htb]
  \centering
  \includegraphics[width=\textwidth]{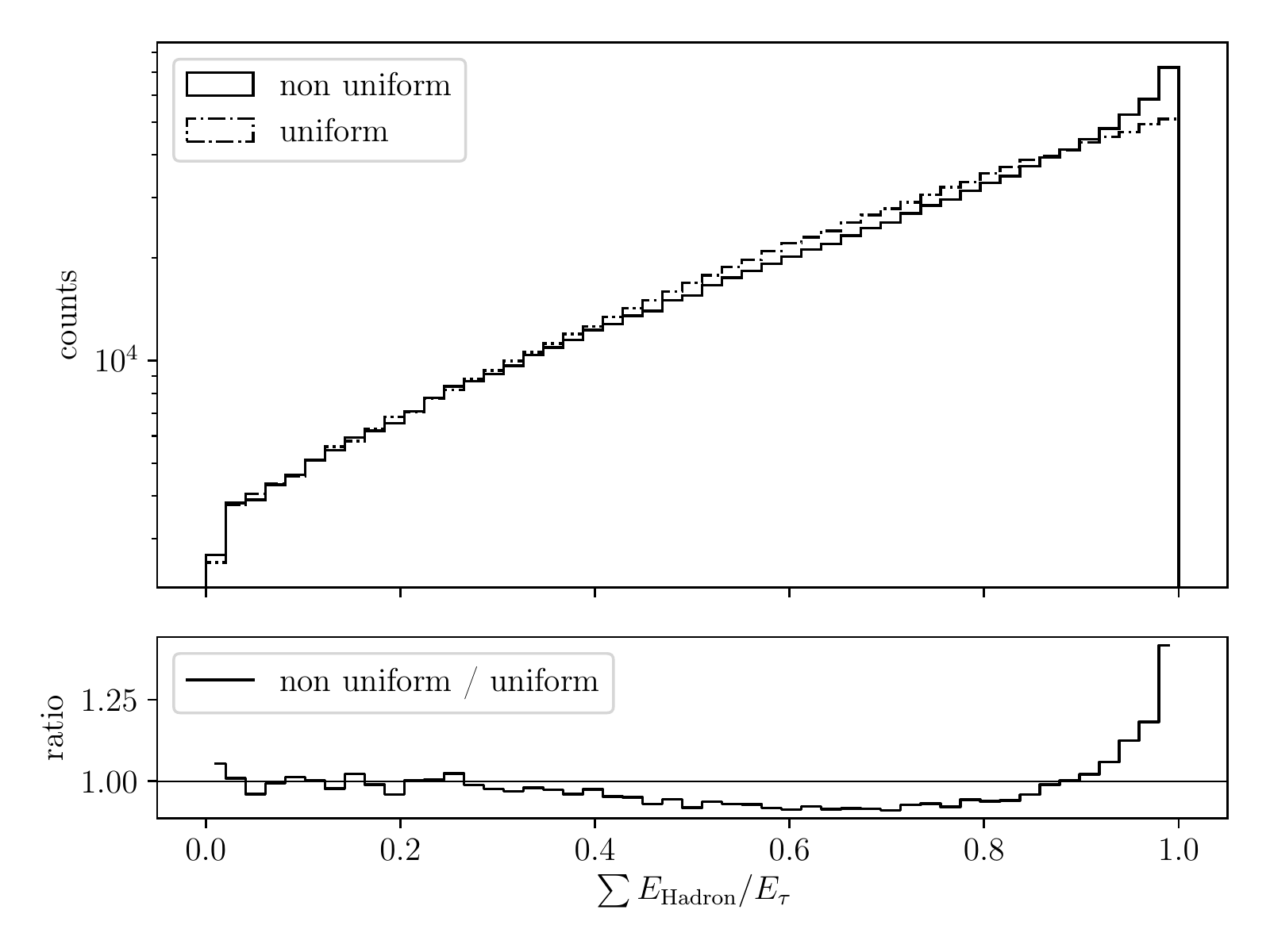}
  \caption{Comparison of the energy distributions of the secondaries for the tau decay with
    non uniform and uniform sampling of events in the phase phase. Shown are the results
    of $10^6$ decays with the decay channels given in \tabref{tab:decay_modes}.}
  \label{fig:tau_decay_compare_uniform_nonuniform}
\end{figure}

\begin{figure}[htb]
  \centering
  \includegraphics[width=\textwidth]{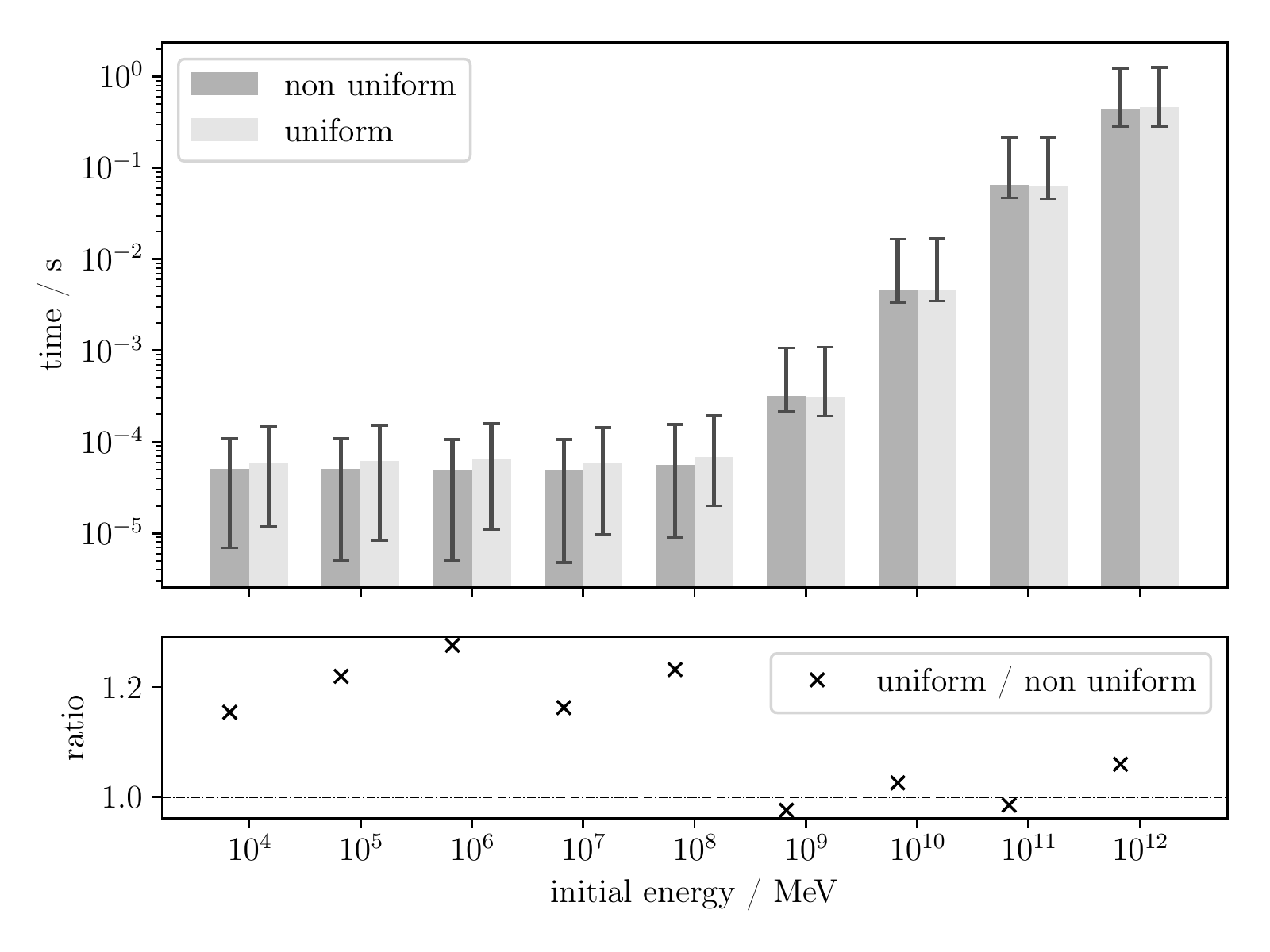}
  \caption{Comparison of run times between the non uniform and uniform phase
    space sampling of the tau decay with the decay channels given in \tabref{tab:decay_modes}.
    Measured are the run times of each tau propagation through ice while per energy $1000$
    taus are propagated.}
  \label{fig:performance_decay}
\end{figure}

With this larger phase space, the energy distribution of the decay modes with
more than two particles is more smooth for the rest and laboratory system as
shown in \figref{fig:decay_dist_n_body}.
This method only improves the phase space sampling; the matrix element is still set to one.
%==============================
%todo may be remove the following lines
Furthermore the Raubold-Lynch algorithm does not sample the momenta uniformly
distributed in the phase space. To create a uniform phase space distribution the rejection
method is used, which results in a performance loss.
The differences of the hadronic energy distributions between uniform and non
uniform sampling are shown in \figref{fig:tau_decay_compare_uniform_nonuniform}.
The performance differences can be seen in
\figref{fig:performance_decay}.

% \begin{figure}[htpb]
%   \centering
%   \includegraphics[width=\textwidth]{plots/decay/tau_decay_compare_uniform_nonuniform_1000000.pdf}
%   \caption{Comparison of the energy distributions of the secondaries for the tau decay with
%     non uniform and uniform sampling of events in the phase phase. Show are the results
%     of $10^6$ decays with the decay channels given in \tabref{tab:decay_modes}.}
%   \label{fig:tau_decay_compare_uniform_nonuniform}
%
%   \vspace*{\floatsep}
%
%   \includegraphics[width=\textwidth]{plots/performance/performance_decay.pdf}
%   \caption{Comparison of run times between the non uniform and uniform phase
%     space sampling of the tau decay with the decay channels given in \tabref{tab:decay_modes}.
%     Measured are the run times of each tau propagation through ice while per energy $1000$
%     taus are propagated.}
%   \label{fig:performance_decay}
% \end{figure}

\begin{figure}[htb]
    \centering
    \includegraphics[width=\textwidth]{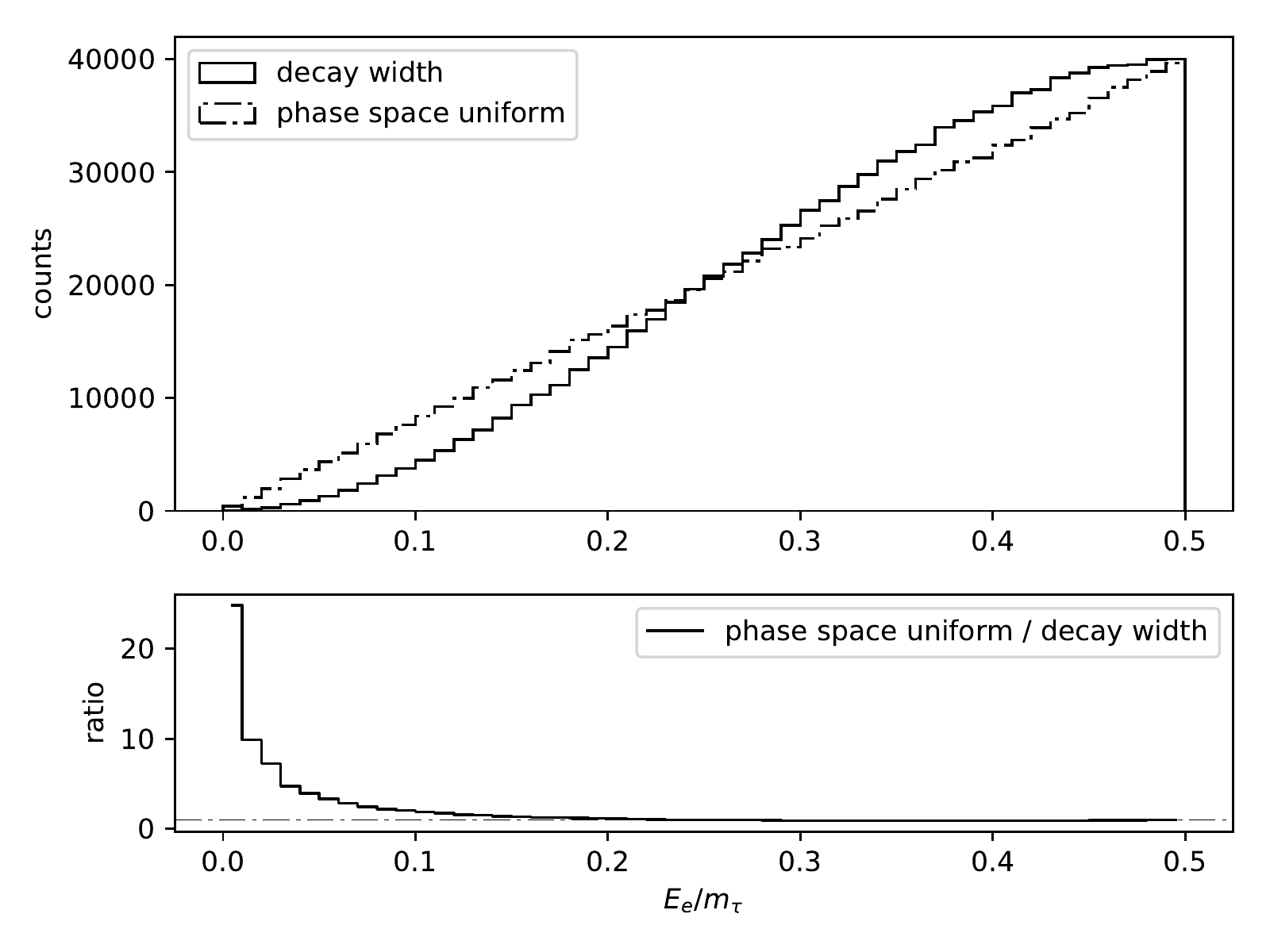}
    \caption{
      Comparison of the final state electron energy distribution of
      $\tau \rightarrow e \nu \nu$ simulated using the differential decay width
      and the uniform phase space sampling.}
    % Comparison of the differential decay width of $\tau \rightarrow e \nu \nu$ to the pure phase space sampling and to the phase space sampling with respect to the matrix elements.}
    \label{fig:decay_dist_compare_uniform}
\end{figure}

\begin{figure}[htb]
    \centering
    \includegraphics[width=\textwidth]{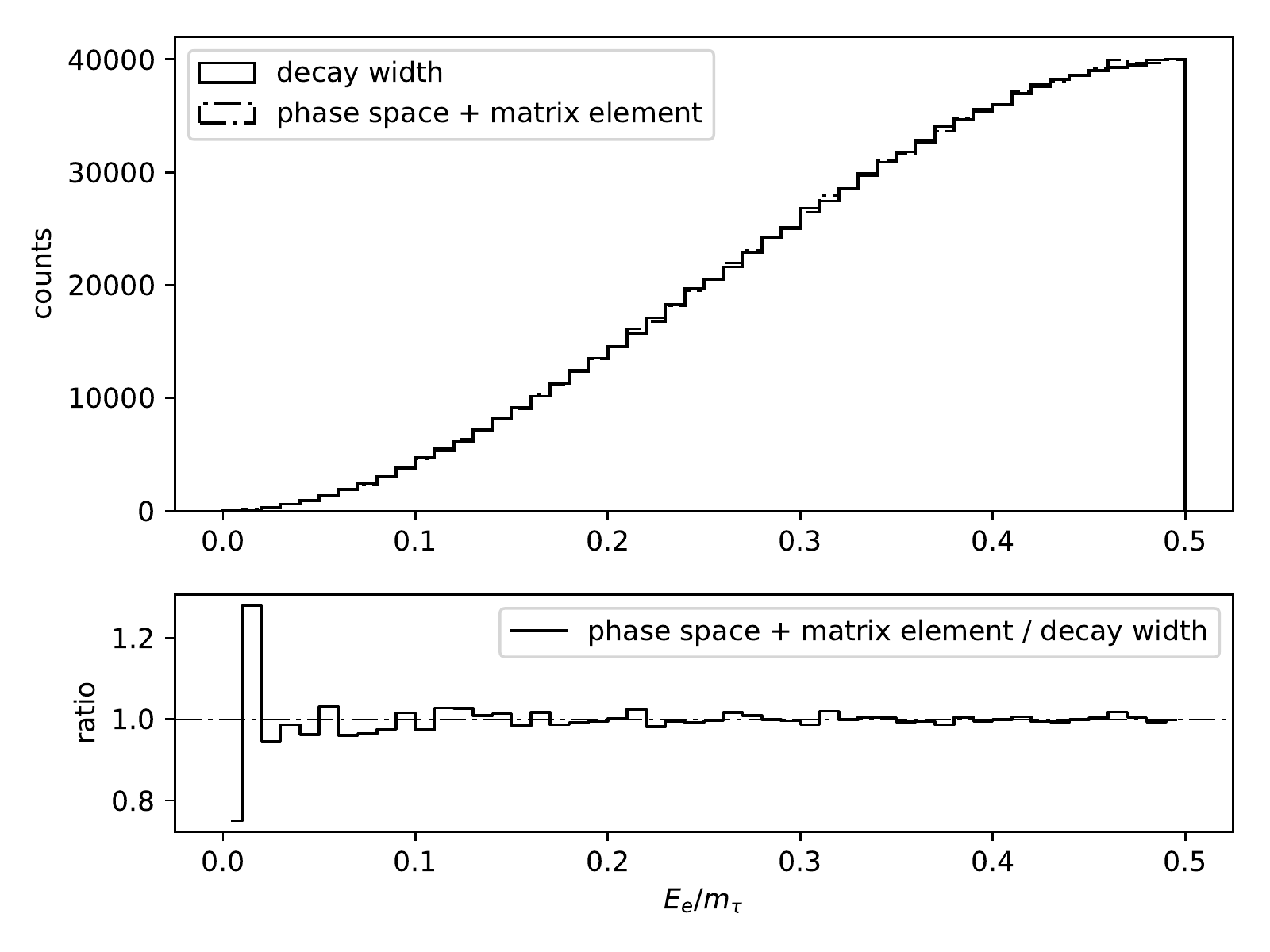}
    \caption{
      Comparison of the final state electron energy distribution of
      $\tau \rightarrow e \nu \nu$ simulated using the differential decay width
      and the phase space sampling with respect to the matrix element.}
      %Comparison of the differential decay width of $\tau \rightarrow \mu \nu
      %\nu$ to the pure phase space sampling and to the phase space sampling
      %with respect to the matrix elements.
    \label{fig:decay_dist_compare}
\end{figure}

%==============================
To illustrate the accuracy of this treatment, the leptonic decay mode with the known energy distribution is compared to the pure phase space calculation.
\figref{fig:decay_dist_compare_uniform} shows the pure phase space sampling of the leptonic decay with constant matrix element compared to the known energy distribution.
If the sampled phase space points are now weighted with the known matrix elements for leptonic decays
\begin{align}
  \mathcal{M} = 64 G_\text{F} \left(p_{\tau} \cdot p_{\nu_l}\right)
                \left(p_{\nu_\tau} \cdot p_{l}\right)
\end{align}
the energy distribution agrees with the differential decay width, as shown in \figref{fig:decay_dist_compare}.

\subsubsection{Multiple Parametrizations for systematic studies}
The systematic uncertainties of the propagation mainly depend on
the uncertainties of the interaction cross section.
Due to the stochastic nature of the processes, a slight shift of one interaction cross section
has great impacts on the probability of the occurrence of the other interactions.
This results in different event signatures which influence the reconstruction
and should be considered in the systematic uncertainties.
Instead of simply shifting the interaction probability
to study the systematic uncertainties,
multiple parametrizations of the cross sections are available,
which changes the probability in a more realistic way.

For bremsstrahlung and inelastic nuclear interaction, multiple
parametrizations already existed in the old version.
For pair production, only the
\cite{KokoulinPetrukhin2, Kelner-atomic} parametrization is implemented.
In the new version, a new parametrization \cite{SSRParam}, shown in \ref{sec:epair_param}, without
the approximation in the structure functions, describing the interaction with the
target atom, is available. This approximation has an uncertainty of around
\SI{3}{\percent}. To further reduce this uncertainty below \SI{1}{\percent},
radiative corrections have to be taken into account.
Additionally, a new Bremsstrahlung
parametrization \cite{SSRParam}, shown in \ref{sec:brems_param}, also without the approximations in the
structure functions and with next to leading order corrections, is now
available.
The effects of these new parametrizations will be illustrated in \cite{SSRParam}.

\begin{figure}[htb]
    \centering
    \includegraphics[width=\textwidth]{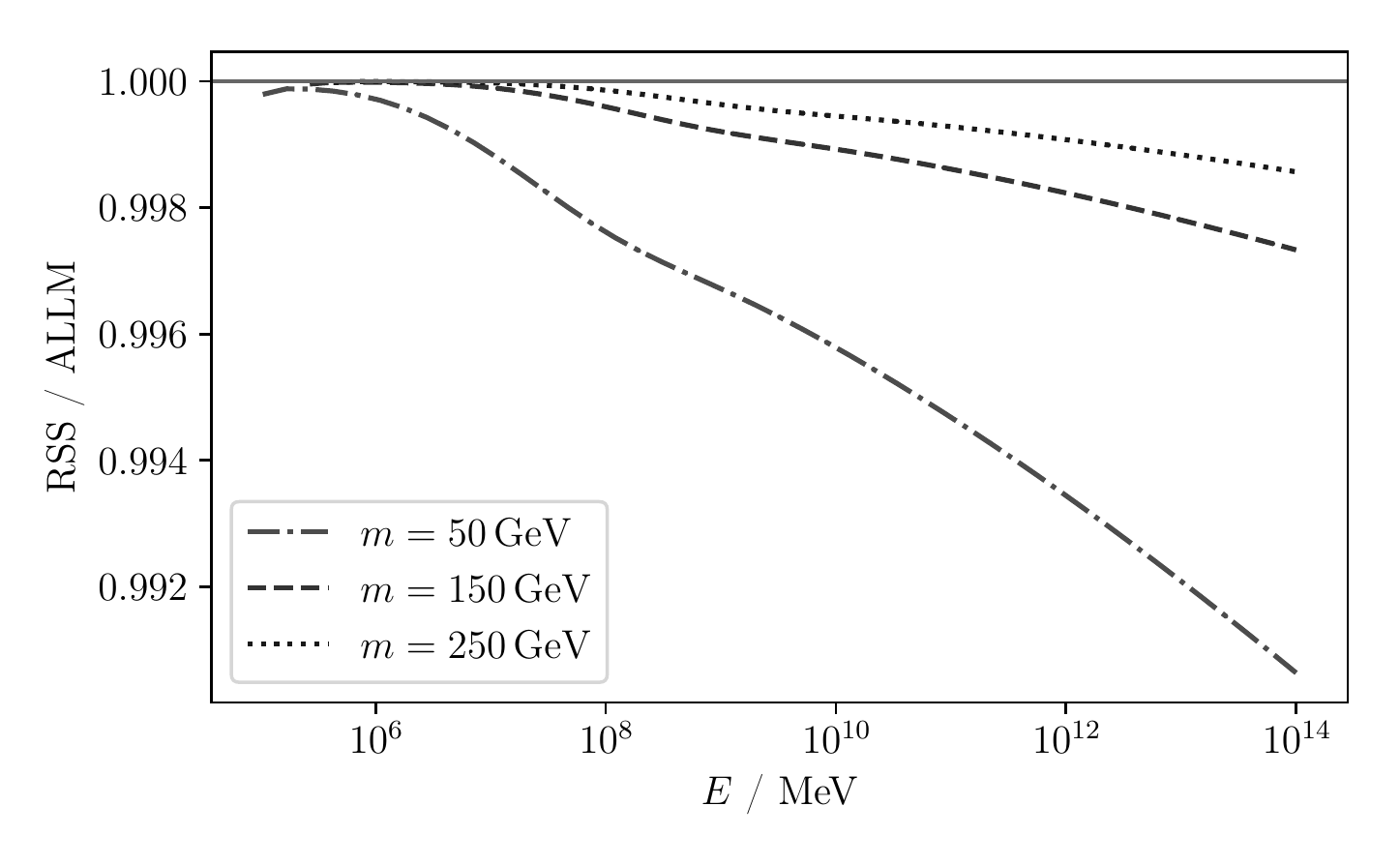}
    \caption{Shown are the energy losses $\rm{d}E/\rm{d}x$ per energy through
      the photonuclear interaction for heavy muons and staus with masses
      $m = \SI{50}{GeV}, \SI{150}{GeV}$ and $\SI{250}{GeV}$. For the muons
      the ALLM97 parametrization is used and for the staus the new RSS
      parametrization.}
    \label{fig:photo_stau}
\end{figure}

Furthermore a new parametrization for the photonuclear interaction is added.
This parametrization describes the interaction of supersymmetric particles,
especially charged sleptons, with nuclei under the exchange of
virtual photons. The derivation of the parametrization is given in \cite{Reno}.
In PROPOSAL this parametrization is called \texttt{PhotoRenoSarcevicSu} (RSS).
\cite{Reno} also shows that this photonuclear interaction is the only
interaction which must be treated differently from the lepton cases.
The reason for adding this parametrization lies in the direct probe of
the supersymmetric breaking scale by finding the supersymmetric particle stau
with the IceCube detector \cite{Albuquerque}.
In earlier analysis, the stau was described as a heavy muon, while using
the photonuclear parametrization of Abramowicz, Levin, Levy and Maor (ALLM97)
\cite{ALLM97}. A comparison of the photonuclear
parametrizations is given in \figref{fig:photo_stau}.

In addition to the new cross section parametrizations, new parametrizations for
multiple scattering, which describes the deviation of the primary lepton to the shower axis, are
implemented. The Highland approximation % (with corrections of Lynch/Dahl)
\cite{Highland75, Lynch91} to the \moliere scattering was implemented, with
considering the decreasing energy during the propagation between two stochastic
losses due to the continuous losses. To accelerate the propagation, a Highland
parametrization with constant energy and no additional integration over the
continuous losses is now available, although this is not a big time consuming
calculation. If the precision of the position is more important than the
simulation time, the original Moli\`{e}re algorithm \cite{Moliere48} can be used.

\begin{figure}[htb]
    \centering
    \includegraphics[width=\textwidth]{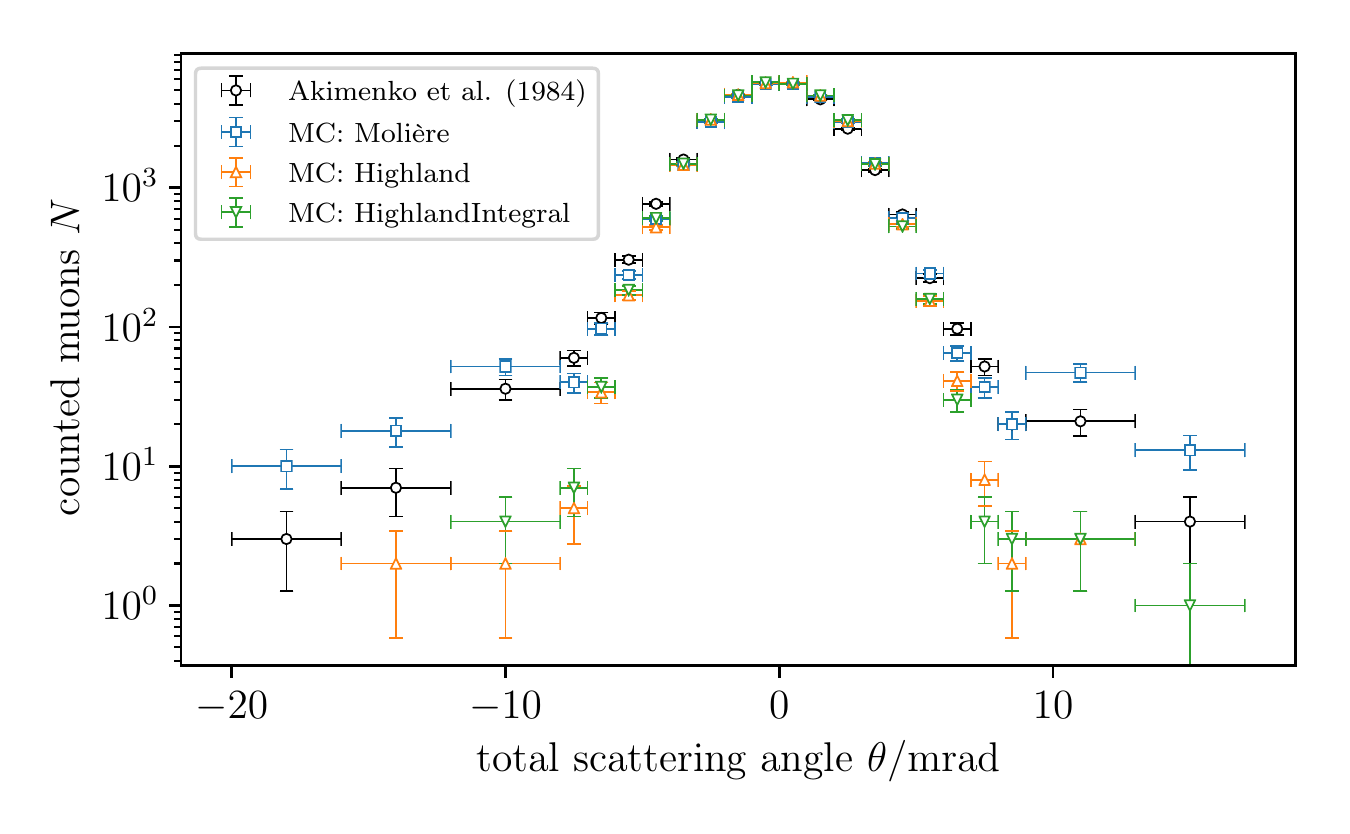}
    \caption{Measured data of $31125$ scattered muons compared with Monte Carlo
      simulations employing the multiple scattering model of Moli\`{e}re, Highland and
      Highland with integration.}
    \label{fig:akimenko}
\end{figure}

\begin{figure}[htb]
    \centering
    \includegraphics[width=0.9\textwidth]{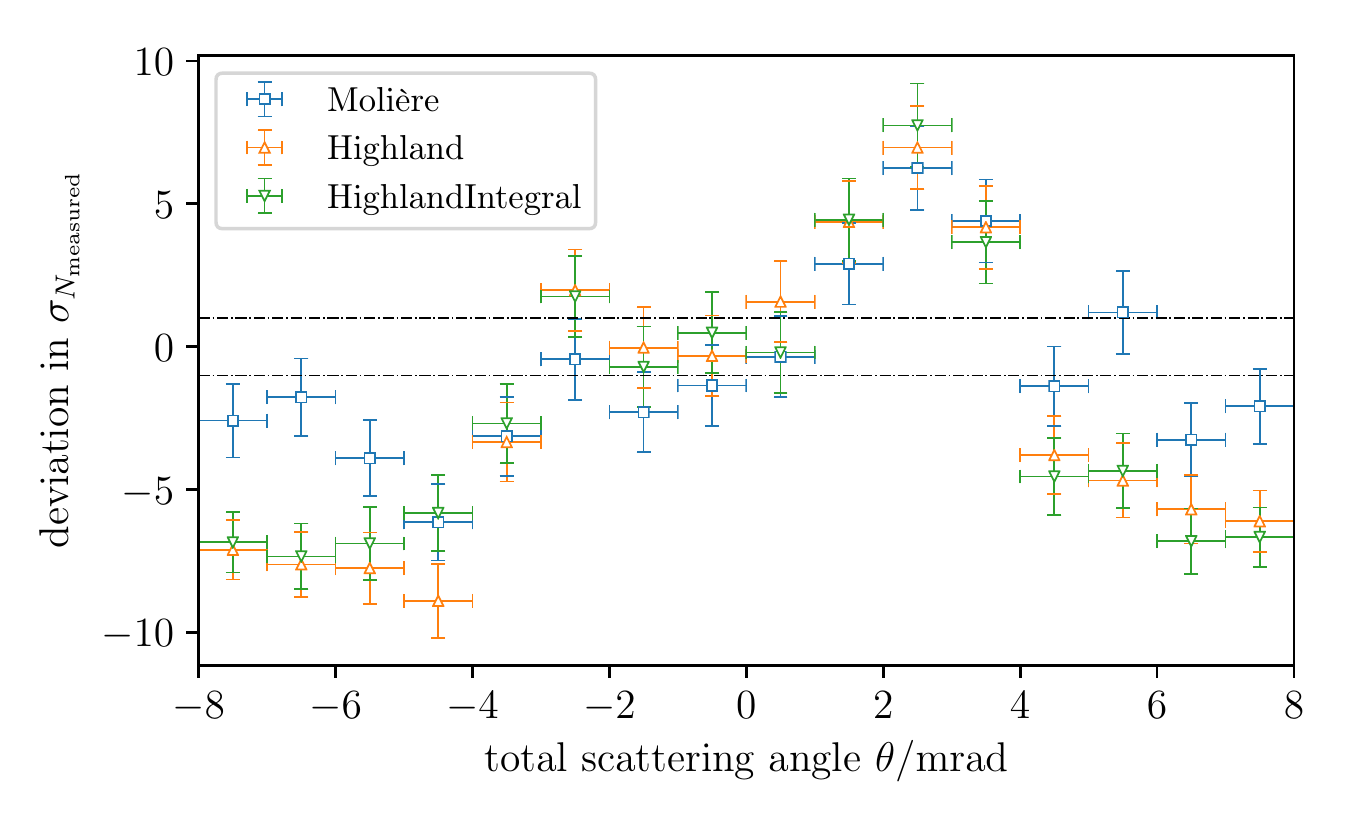}
    \caption{Deviation of the Monte Carlo data from the measurement in terms
      of the Poisson error of the measured data $\sigma_{N_\text{measured}}$.}
    \label{fig:scattering_ratio}
\end{figure}

To validate the implemented multiple scattering models, measured data of Akimenko et.~al
\cite{akimenko} is used. They measured the deviation of muons with
a momentum of \SI{7.3}{\GeV\per\speedoflight} traversing \SI{1.44}{cm}
($\approx 1$ radiation length) of copper.
Such muons with all three implemented multiple scattering models were simulated.
The comparison of the projected scattering angles $\theta$
with the data and the Monte Carlo simulation can be seen in \figref{fig:akimenko}.
The deviations of the Monte Carlo data to
the measured data is shown in \figref{fig:scattering_ratio}.
For this scenario the RMS of the Highland approximation is
$\theta_0 = \SI{1.863}{\milli\rad}$. The deviation plot in \figref{fig:scattering_ratio}
shows, that the \moliere model is in good agreement with the data up to
$4.5 \cdot \theta_0$ whereas the Highland models are no longer acceptable when they 
exceed $2 \cdot \theta_0$. In addition, \figref{fig:akimenko} shows
that for larger scattering angles the \moliere model overestimates the
measured data while both Highland models underestimate them.

\begin{figure}[htb]
    \centering
    \includegraphics[width=0.9\textwidth]{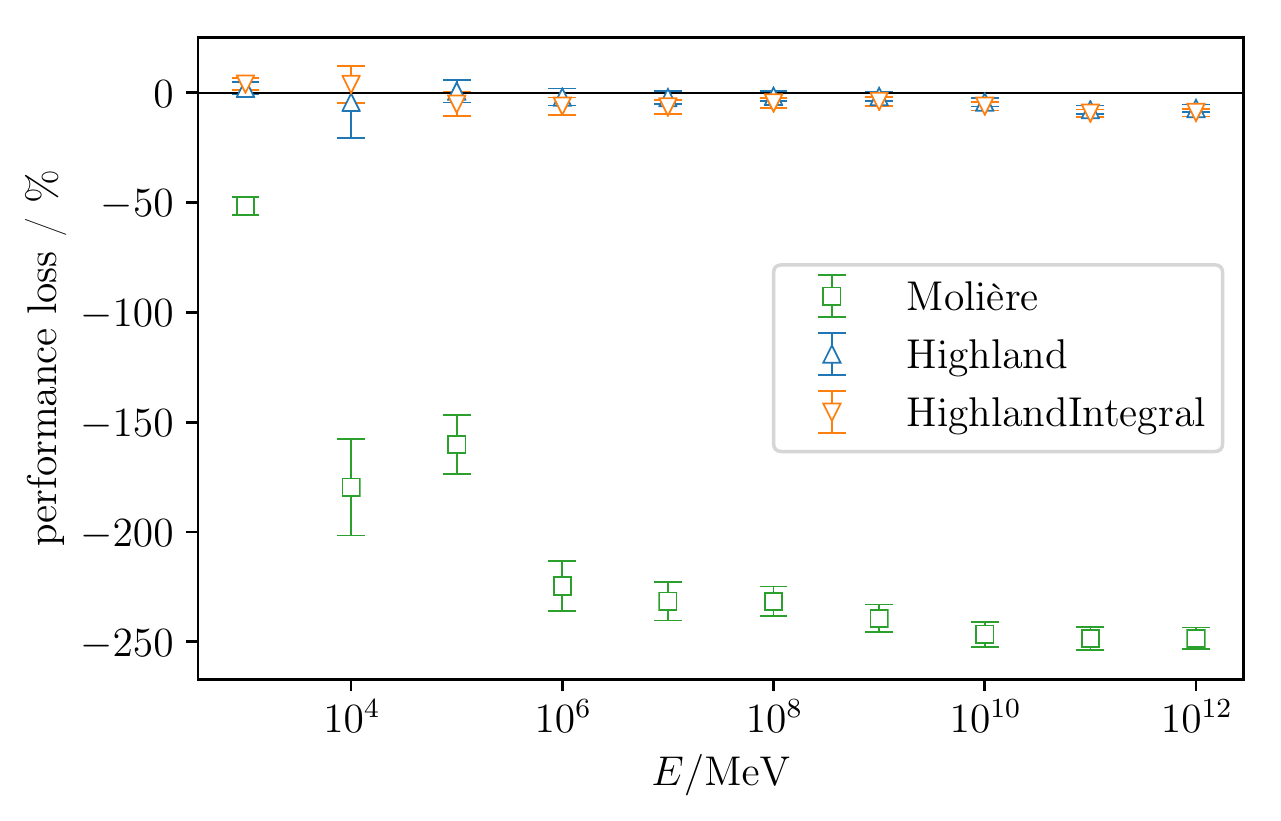}
    \caption{Performance loss of the implemented multiple scattering models compared
    to disabled multiple scattering. At each energy $10^5$ muons are propagated
    \SI{100}{m} through \frejus rock.}
    \label{fig:performance_all_frejus}
\end{figure}

\begin{figure}[htb]
    \centering
    \includegraphics[width=0.85\textwidth]{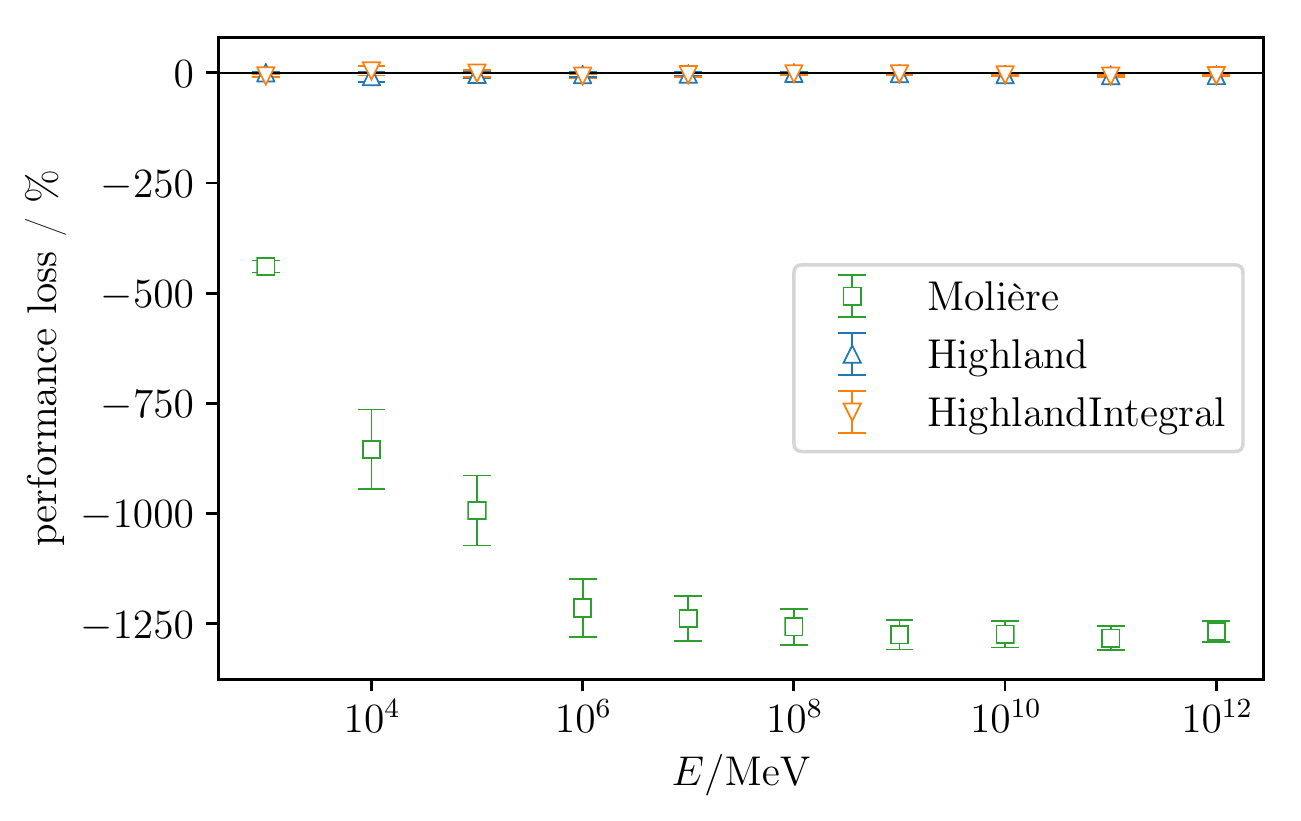}
    \caption{Performance loss of the implemented multiple scattering models compared
    to disabled multiple scattering. At each energy $10^5$ muons are propagated
    \SI{100}{m} through ANTARES water \cite{antares}.}
    \label{fig:performance_all_antares}
\end{figure}

\begin{figure}[htb]
    \centering
    \includegraphics[width=0.85\textwidth]{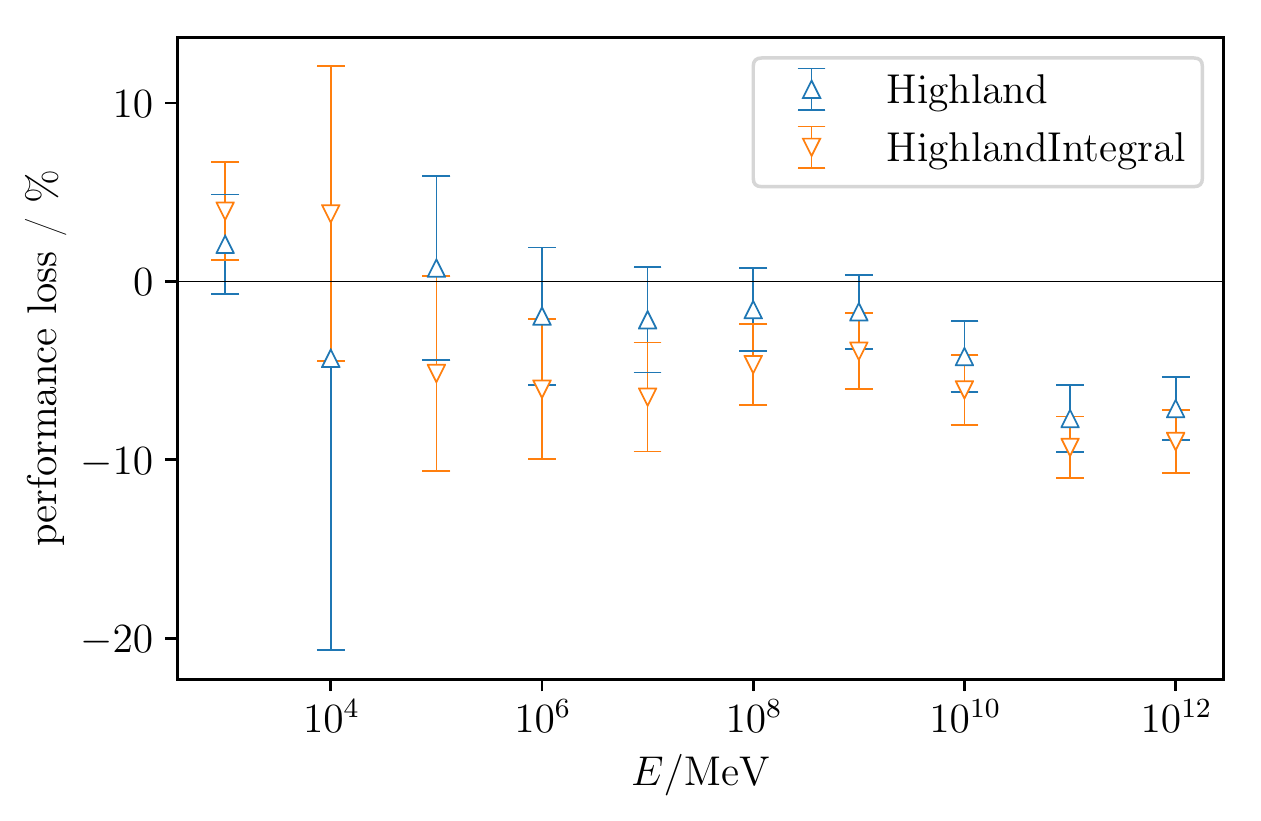}
    \caption{Performance loss of the implemented Highland and
      HighlandIntegral model compared
    to disabled multiple scattering. At each energy $10^5$ muons are propagated
    \SI{100}{m} through \frejus rock.}
    \label{fig:performance_all_HHI}
\end{figure}

Although the \moliere model gives more reliable results, the performance costs
are worth noting. In \figref{fig:performance_all_frejus} the performance losses
of all three multiple scattering models are presented compared to disabled multiple scattering.
It can be seen that the propagation with the \moliere scattering takes
up to \SI{250}{\%} longer, especially for higher energies, while the
performance loss for both Highland models is almost negligible.
This situation is even worse when choosing a medium with more components
since the \moliere model is extremely sensitive to the number of components.
\figref{fig:performance_all_antares} shows the same setup as
\figref{fig:performance_all_frejus} except for choosing ANTARES water \cite{antares}
as the medium. In PROPOSAL ANTARES water is implemented with $8$ different components compared
to \frejus rock with technically one component.
This causes a performance loss of up to a factor of \num{12} for the propagation with
\moliere scattering enabled (\figref{fig:performance_all_antares}).
A comparison of the Highland models is given in \figref{fig:performance_all_HHI}.
Since the integration of the original HighlandIntegral model is removed for
the Highland model, the propagation with Highland scattering is slightly
faster.
The time measurements for this section have been performed on a laptop
computer with an \textit{Intel\textsuperscript{~\textregistered} Core\textsuperscript{\texttrademark} i5-4200U}
processor.


% \begin{figure}
%     \centering
%     \includegraphics[width=\textwidth]{plots/scat_time.pdf}
%     \caption{Average time to propagate muons with different scattering parametrizations.}
%     \label{fig:scat_time}
% \end{figure}

\clearpage
\section{Conclusion}
\label{sec:conclusion}

A new version of the lepton propagator PROPOSAL is presented. Compared to the
previous version, computational as well as physical improvements are achieved.
This new version of PROPOSAL is used in the simulation chain of the IceCube
detector.

The description of particles was changed following a polymorphism paradigm.
This allows one to add new particles and change the properties of already
implemented particles before initialization, which is useful to investigate physics beyond
the standard model and for systematic studies. The stau was added as an
example of this new possibility.

In addition more recent cross sections for the energy loss processes of muons were added,
and the description of hadronic tau decays was improved by dropping the two-particle
decay approximation used in the previous version. This leads to more realistic
secondary particle spectra.

The polymorphism and other improvements resulted in a performance increase of
about \SI{30}{\%} compared to the previous version. This already includes the
slight loss of speed due to the more exact treatment of tau decays etc.

\section*{Acknowledgments}
We acknowledge funding by the Helmholtz-Allianz für Astroteilchenphysik 
under the grant number HA-301 % (A.~S, W.~R.)
and by the Deutsche Forschungsgemeinschaft under the grant number RH 25/9-1. % (J.~S., W.~R.).
This work has been supported by the DFG, Collaborative Research Center SFB 876,
project C3 (http://sfb876.tu-dortmund.de).
We thank the IceCube Collaboration for numerous useful discussions,
in particular Alexander Olivas, Jakob van Santen and Christopher Weaver.

The authors also thank Tomasz Fuchs, Malte Geiselbrinck and Jan-Hendrik Köhne for
useful discussions. We are grateful to Anthony Flores for careful proofreading.

%\linenumbers
% \section*{References}
\bibliography{references}
\appendix
\section{Improved cross section parametrizations}
The cross section parametrizations reported here will be discussed in detail
in a separate publication \cite{SSRParam}.

In this appendix, following symbols are used:
\begin{center}
\begin{tabular}{cp{0.7\textwidth}}
$B$ & radiation logarithm ($\approx 183$) \cite{KKP,RadiationLogarithm}\\
$B'$ & inelastic radiation logarithm ($\approx 1429$) \cite{Kelner-atomic}\\
$D_n = 1.54 A^{0.27}$ & nuclear formfactor parametrization \cite{KKP}\\
$\mu$ & mass of the incoming particle\\
\end{tabular}
\end{center}

\subsection{Bremsstrahlung} \label{sec:brems_param}
This parametrization takes into account: elastic atomic and nuclear formfactors,
inelastic nuclear formfactors, bremsstrahlung on atomic electrons ($\mu$-%
diagrams only), and radiative corrections.
\begin{equation}
\begin{split}
\frac{d\sigma}{dv} &= 4 Z^2 \alpha \left(r_e \frac{m_e}{\mu}\right)^2
  \frac{1}{v} \left\{
  \left[(2 - 2 v + v^2) \Phi_1(\delta) - \frac{2}{3} (1 - v) \Phi_2(\delta)
  \right] \right.\\
&+ \left.\frac{1}{Z} s_\text{atomic}(v, \delta) + \frac{\alpha}4 \Phi_1(\delta)
  s_\text{rad}(v) \right\},
\end{split}
\end{equation}
where
\begin{align}
\Phi_1(\delta) &= \ln \frac{\frac{\mu}{m_e} B Z^{-1/3}}{1 + B Z^{-1/3} \sqrt{e}
  \delta/m_e} - \Delta_1 \left(1 - \frac{1}{Z}\right),\\
\Phi_2(\delta) &= \ln \frac{\frac{\mu}{m_e} B Z^{-1/3} e^{-1/6}}{1 + B Z^{-1/3}
  e^{1/3} \delta/m_e} - \Delta_2 \left(1 - \frac{1}{Z}\right),\\
\Delta_1 &= \ln\frac{\mu}{q_c} + \frac{\rho}{2} \ln \frac{\rho + 1}{\rho - 1},\\
\Delta_2 &= \ln\frac{\mu}{q_c} + \frac{3 \rho - \rho^3}{4} \ln\frac{\rho + 1}
  {\rho - 1} + \frac{2 \mu^2}{q_c^2},\\
\rho &= \sqrt{1 + \frac{4 \mu^2}{q_c^2}}, q_c = m_\mu e/D_n,\\
s_{\text{atomic}}(\delta) &= \left[\frac{4}{3}(1 - v) + v^2\right] \left[
  \ln \frac{\mu/\delta}{\mu \delta/m_e^2 + \sqrt{e}} - \ln \left(1
  + \frac{m_e}{\delta B' Z^{-2/3} \sqrt{e}}\right) \right],\\
s_\text{rad}(v) &= \begin{cases}
\sum_{n = 0}^2 a_n v^n & v < 0.02,\\
\sum_{n = 0}^3 b_n v^n & 0.02 \leq v < 0.1,\\
\sum_{n = 0}^2 c_n v^n + c_3 v \ln v + c_4 \ln(1 - v) + c_5 \ln^2(1 - v) & 0.1
  \leq v < 0.9,\\
\sum_{n = 0}^2 d_n v^n + d_3 v \ln v + d_4 \ln(1 - v) + d_5 \ln^2(1 - v) & v
  \geq 0.9,
\end{cases}
\end{align}
where the values of the fit parameters $a_n, b_n, c_n, d_n$ are given in table~%
\ref{tab:rad_params}.
\begin{table}
\begin{tabular}{crrrrrr}
\toprule
$n$ & 0 & 1 & 2 & 3 & 4 & 5\\
\midrule
$a_n$ & $-$0.00349 & 148.84 & $-$987.531\\
$b_n$ & 0.1642 & 132.573 & $-$585.361 & 1407.77\\
$c_n$ & $-$2.8922 & $-$19.0156 & 57.698 & $-$63.418 & 14.1166 & 1.84206\\
$d_n$ & 2134.19 & 581.823 & $-$2708.85 & 4767.05 & 1.52918 & 0.361933\\
\bottomrule
\end{tabular}
\caption{Parameters of the parametrization for the radiative corrections to the
  bremsstrahlung cross section.}
\label{tab:rad_params}
\end{table}

\subsection{Pair production} \label{sec:epair_param}
This parametrization of the pair production cross section takes into account:
elastic atomic and nuclear formfactors, and pair
production on atomic electrons\footnote{Because of the way this calculation is set up, it
is impossible to take into account the inelastic nuclear formfactor and the
atomic electron contribution simultaneously.}.
\begin{equation}
\frac{d^2\sigma}{dv\,d\rho} = \frac{2}{3\pi} Z (Z + \zeta) \frac{1 - v}{v}
  \left[\Phi_e + \frac{m_e^2}{m_\mu^2} \Phi_\mu\right],
\end{equation}
where 
\begin{align}
\Phi_e &= C_1^e L_1^e + C_2^e L_e^2,\\
C_1^e &= C_e - C_2^e,\\
C_2^e &= [(1 - \rho^2)(1 + \beta) + \xi (3 - \rho^2)] \ln\left(1 + \frac{1}
  {\xi}\right) + 2 \frac{1 - \beta - \rho^2}{1 + \xi} - (3 - \rho^2),\\
L_1^e &= \ln \frac{B Z^{-1/3} \sqrt{1 + \xi}}{X_e + \frac{2 m_e \sqrt{e} B
  Z^{-1/3} (1 + \xi}{E v (1 - \rho^2}} - \frac{\Delta_e}{C_e}
  - \frac{1}{2} \ln \left[X_e + \left(\frac{m_e}{m_\mu} D_n\right)^2 (1 + \xi)
  \right]\\
L_2^e &= \ln \frac{B Z^{-1/3} e^{-1/6}\sqrt{1 + \xi}}{X_e + \frac{2 m_e
  e^{1/3} B Z^{-1/3} (1 + \xi}{E v (1 - \rho^2)}} - \frac{\Delta_e}{C_e}
  - \frac{1}{2} \ln \left[X_e + \left(\frac{m_e}{m_\mu} D_n\right)^2 e^{1/3}
  (1 + \xi) \right]\\
X_e &= \exp\left(-\frac{\Delta_e}{C_e}\right),\\
C_e &= [(2 + \rho^2) (1 + \beta) + \xi (3 + \rho^2)] \ln\left(1 +
  \frac{1}{\xi}\right) + \frac{1 - \rho^2 - \beta}{1 + \xi} - (3 + \rho^2),\\
\Delta_e &= [(2 + \rho^2)(1 + \beta) + \xi (3 + \rho^2)] \Li_2 \frac{1}{1 + \xi}
  - (2 + \rho^2) \xi \ln \left(1 + \frac{1}{\xi}\right) - \frac{\xi + \rho^2
  + \beta}{1 + \xi},
\end{align}
where $L_{1,2}^e$ can be equivalently expressed in the case of large $X_e$ as
\begin{align}
L_1^e &= \ln \frac{B Z^{-1/3} \sqrt{1 + \xi}}{1 + \frac{2 m_e \sqrt{e} B Z^{-1/3}
  (1 + \xi)}{E v (1 - \rho^2)} X_e^{-1}} - \frac{1}{2} \frac{\Delta_e}{C_e}
  - \frac{1}{2} \ln \left[1 + \left(\frac{m_e}{m_\mu} D_n\right)^2 (1 + \xi)
  X_e^{-1}\right],\\
L_2^e &= \ln \frac{B Z^{-1/3} e^{-1/6} \sqrt{1 + \xi}}{1 + \frac{2 m_e e^{1/3} B Z^{-1/3}
  (1 + \xi)}{E v (1 - \rho^2)} X_e^{-1}} - \frac{1}{2} \frac{\Delta_e}{C_e}
  - \frac{1}{2} \ln \left[1 + \left(\frac{m_e}{m_\mu} D_n\right)^2 e^{1/3}(1 + \xi)
  X_e^{-1}\right],
\end{align}
and
\begin{align}
\Phi_\mu &= C_1^\mu L_1^\mu + C_2^\mu L_2^\mu,\\
L_1^\mu &= \ln\frac{B \frac{\mu}{m_e} Z^{-1/3}/D_n}{X_\mu + \frac{2 m_e \sqrt{e}
  B Z^{-1/3} (1 + \xi)}{E v (1 - \rho^2)}} - \frac{\Delta_\mu}{C_\mu},\\
L_2^\mu &= \ln\frac{B \frac{\mu}{m_e} Z^{-1/3}/D_n}{X_\mu + \frac{2 m_e e^{1/3}
  B Z^{-1/3} (1 + \xi)}{E v (1 - \rho^2)}} - \frac{\Delta_\mu}{C_\mu},\\
C_1^\mu &=  C_\mu - C_2^\mu,\\
C_2^\mu &= [(1 - \beta)(1 - \rho^2) - \xi (1 + \rho^2)] \frac{\ln (1 + \xi)}
  {\xi} - 2 \frac{1 - \beta - \rho^2}{1 + \xi} + 1 - \beta - (1 + \beta)
  \rho^2,\\
C_\mu &= \left[(1 + \rho^2) \left(1 + \frac{3}{2} \beta\right) - \frac{1}{\xi}
  (1 + 2 \beta) (1 - \rho^2)\right] \ln (1 + \xi) \\
  &+ \frac{\xi (1 - \rho^2
  - \beta)}{1 + \xi} + (1 + 2 \beta) (1 - \rho^2),\\
X_\mu &= \exp\left(-\frac{\Delta_\mu}{C_\mu}\right),\\
\begin{split}
\Delta_\mu &= \left[(1 + \rho^2) \left(1 + \frac{3}{2} \beta\right)
  - \frac{1}{\xi} (1 + 2 \beta) (1 - \rho^2)\right] \Li_2\left(\frac{\xi}{1
  + \xi}\right)\\ &+ \left(1 + \frac{3}{2} \beta\right) \frac{1 - \rho^2}{\xi}
  \ln (1 + \xi) + \left[1 - \rho^2 - \frac{\beta}{2} (1 + \rho^2) + 
  \frac{1 - \rho^2}{2 \xi} \beta\right] \frac{\xi}{1 + \xi},
\end{split}
\end{align}
where $L_{1,2}^\mu$ can be expressed for large $X_\mu$ equivalently as
\begin{align}
L_1^\mu &= \ln\frac{B \frac{\mu}{m_e} Z^{-1/3}/D_n}{1 + \frac{2 m_e \sqrt{e}
  B Z^{-1/3} (1 + \xi)}{E v (1 - \rho^2)} X_\mu^{-1}},\\
L_2^\mu &= \ln\frac{B \frac{\mu}{m_e} Z^{-1/3}/D_n}{1 + \frac{2 m_e e^{1/3}
  B Z^{-1/3} (1 + \xi)}{E v (1 - \rho^2)} X_\mu^{-1}},
\end{align}
with the abbreviations
\begin{align}
\beta &= \frac{v^2}{2 (1 - v)},\\
\xi &= \left(\frac{\mu v}{m_e}\right)^2 \frac{1 - \rho^2}{1 - v},\\
\zeta &= \frac{0.073 \ln \frac{E/\mu}{1 + \gamma_1 Z^{2/3} E/\mu} - 0.26}
  {0.058 \ln \frac{E/\mu}{1 + \gamma_2 Z^{1/3} E/\mu} - 0.14},\\
\gamma_1 &= \num{1.95e-5}, \quad \gamma_2 = \num{5.3e-5} \text{ for } Z \neq 1,
  \\
\gamma_1 &= \num{4.4e-5}, \quad \gamma_2 = \num{4.8e-5} \text{ for } Z = 1.
\end{align}
The dilogarithm $\Li_2(x)$ is defined as
\begin{equation}
\Li_2(x) = -\operatorname{Re} \int_0^x \frac{\ln(1 - t)}{t} dt.
\end{equation}

\end{document}